\documentclass[12pt,preprint]{aastex}
\shorttitle{\sc STIS UV Spectra of Nearby Star-forming Regions}
\shortauthors{\sc Schwartz et al.}

\def\kms{km~s$^{-1}$}
\def\ha{H$\alpha$~}
\def\luv{L$_{UV}$}

\begin{document}

%__________________________________________________________________________

\title{Kinematics of Interstellar Gas in Nearby UV-Selected
Galaxies Measured with {\em HST}/STIS Spectroscopy\altaffilmark{1} }

\author{C. M. Schwartz, C. L. Martin\altaffilmark{2,3}} 
\affil{Department of Physics, University of California, Santa Barbara,
CA 93106, colleen@physics.ucsb.edu, cmartin@physics.ucsb.edu}

\author{R. Chandar, C. Leitherer}
\affil{Space Telescope Science Institute, 3700 San Martin Drive,
  Baltimore, MD 21218}

\author{T. M. Heckman}
\affil{Department of Physics and Astronomy, Johns Hopkins University,
  3400 North Charles Street, Baltimore, MD 21218}

\author{M. S. Oey}
\affil{Department of Astronomy, 830 Dennison Building, University of
  Michigan, Ann Arbor, MI 48109} 

\altaffiltext{1}{Based on observations made with the NASA/ESA {\em
  Hubble Space Telescope}, obtained from the Data Archive at the Space
  Telescope Science Institute, which is operated by the Association of
  Universities for Research in Astronomy, Inc., under NASA contract
  NAS 5-26555. These observations are associated with program GO-9036.}

\altaffiltext{2}{Packard Fellow}
\altaffiltext{3}{Alfred P. Sloan Foundation Fellow}

%__________________________________________________________________________

\begin{abstract}
We measure Doppler shifts of interstellar absorption lines in {\em
HST}/STIS spectra of individual star clusters in nearby
UV-selected galaxies.  Values for systemic velocities, which are needed to 
quantify outflow speeds, are taken from the literature, and verified with 
stellar lines.  We detect outflowing gas in eight of 17 galaxies
via low-ionization lines (e.g., \ion{C}{2}, \ion{Si}{2}, \ion{Al}{2}), 
which trace cold and/or warm gas.  The starbursts in our sample are
intermediate in luminosity (and mass) to dwarf galaxies and luminous
infrared galaxies (LIRGs), and we confirm that their outflow speeds
(ranging from $-100$~\kms~to nearly $-520$~\kms~with an accuracy of
$\sim$80~\kms) are intermediate to those previously measured in dwarf
starbursts and LIRGs.  We do not detect the outflow in high-ionization lines
(such as \ion{C}{4} or \ion{Si}{4}); higher quality data will be
needed to empirically establish how velocities vary with the
ionization state of the outflow. We do verify that the low-ionization
UV lines and optical \ion{Na}{1} doublet give roughly consistent
outflow velocities solidifying an important link between studies of 
galactic winds at low and high redshift. To obtain higher
signal-to-noise, we create a local average composite spectrum, and
compare it to the high-$z$ Lyman Break composite spectrum.
Surprisingly, the low-ionization lines show similar outflow velocities
in the two samples.  We attribute this to a combination of weighting
towards higher luminosities in the local composite, as well as both
samples being on average brighter than the ``turnover'' luminosity in
the $v - SFR$ relation.
\end{abstract}

%The low-ionization interstellar absorption lines in our average
%sample composite spectrum show a blueshift similar to that of a
%composite spectrum for high-redshift Lyman-break-selected galaxies,
%but are broader due to a difference in spectral resolution.  
%__________________________________________________________________________

\keywords{galaxies: ISM -- galaxies: kinematics and dynamics --
  galaxies: starburst -- ISM: outflows}  

%__________________________________________________________________________

\section{Introduction\label{intro}}

Galaxy -- galaxy encounters likely trigger a significant fraction 
of all galactic star formation (e.g., Kennicutt 1989; Heckman
1998). The strongest starburst events often produce galactic-scale
winds, which transport metals and dust from the interstellar medium
(ISM) into the galactic halo, and possibly beyond (e.g., Larson 1974;
Dekel \& Silk 1988, Lehnert \& Heckman 1996).  These outflows are
thought to heat and enrich the intergalactic medium (IGM) and even the
intracluster medium. Although the amount of matter permanently
escaping the galaxy is still under debate, the interplay between star
formation and galactic winds clearly plays a fundamental role in the
evolution of galaxies. While feedback from active galactic nuclei may
be important in larger  galaxies  (Scannapieco \& Oh 2004),  the
deposition of energy and dispersal of metals by supernovae appear to
shape the properties of smaller galaxies.  In this paper, we explore
the relation between outflow properties and galaxy morphology,
environment, and star formation history via ultraviolet spectroscopy
of local star-forming galaxies.    

Understanding the large-scale movements of material from the ISM into the
IGM requires a panchromatic picture because the temperature of the
outflowing gas spans a broad temperature range. Winds were first
recognized in emission lines from warm, photoionized gas.  Recent
studies have observed cold gas via UV/optical resonance absorption
lines (Heckman et al. 2000, Rupke et al. 2002, Schwartz \& Martin
2004, Martin 2005). The warm and cold gas is thought to be swept-up by
a much hotter wind (and/or entrained in it). X-ray observations detect
the portion of the hot wind  around $0.8$~keV and find that this hot
gas contains the metals from the supernova ejecta  (Dahlem et al. 1998;
Martin et al. 2002).

Much recent work has focused on absorption line studies of winds. 
Using the starburst itself as a background continuum source, the
Doppler shifts of lines determine absolutely whether the gas 
is falling in (redshifted) or expanding outward
(blueshifted). Moreover, unlike emission lines, whose strength scales
as density squared, the strength of absorption lines are directly
proportional to the column density of the ion  along the
sightline. Measurements for both local and distant starbursts often
reveal blueshifted, interstellar absorption lines.  For example,
Schwartz \& Martin (2004) measure outflows in nearby dwarf starbursts
at speeds around $-30$~\kms.  Heckman et al. (2000) find an average
outflow speed of $-100$~\kms~for luminous infrared galaxies (LIRGs).
Martin (2005) and Rupke et al. (2002) detect outflow speeds of up to
$-700$~\kms~for two samples of ultraluminous infrared galaxies (ULIRGs). 
When these data for local starbursts are combined, the maximum outflow 
speeds measured in low-ionization lines are seen to increase %fairly quickly
with the global star formation rate (Martin 2005). The outflow speeds
measured in distant starbursts (e.g. Shapley et al. 2003;  Pettini et
al. 2002; Swinbank et al. 2004), redshifts $z \sim 3$, appear to be  
consistent with this empirical velocity -- SFR relation.
However, this comparison implicitly assumes that \ion{Na}{1} absorption
(rest-frame optical) is tracing the same outflow structures as the
low-ionization UV lines measured for the high-redshift galaxies. 

Studying the kinematics of ultraviolet, interstellar lines  in
nearby galaxies serves several purposes. It provides a direct
comparison between velocities measured from \ion{Na}{1} and UV
lines. It also increases the number of intermediate-luminosity
starbursts with measured outflow velocities, which is important
because the velocities from just three dwarf starbursts 
(Schwartz \& Martin 2004) provide a lot of the leverage in the  
$v - SFR$ relation (Martin 2005). Additionally, the rest-frame
ultraviolet (UV) spectrum includes a wealth of resonance lines that
sample a broader range of ionization states than do optical resonance
lines (e.g., Kinney et al. 1993; Heckman \& Leitherer 1997; Heckman et
al. 1998; Kunth et al. 1998; Shapley et al. 2003; Chandar et al. 2004;
V\'azquez et al. 2004).  In principle, measurements of ultraviolet
resonance lines can determine how the outflow speed varies among
different temperature components of the wind.  For example,  V\'azquez
et al. 2004 suggest an increase in outflow velocity with ionization
energy in a high-resolution spectrum of a local dwarf starburst,
NGC~1705. 

The UV cannot be observed in nearby galaxies from ground-based telescopes,
so the spectroscopic capabilities of the Hubble Space Telescope provide
unique data. As part of GO Program \#9036, we examine UV resonance
lines in Space Telescope Imaging Spectrograph (STIS) data of 17 nearby
UV-selected galaxies. The least blended interstellar lines, which best
constrain the gas kinematics, have first ionization energies less
than that of H~I. Hence, we obtain  the most information about the
velocity of {\it cold} (i.e., $\sim 10^2$~K)  and/or {\it warm}
($\lesssim 10^4$~K) interstellar gas, where much of the  hydrogen may
be neutral.  In principle, hotter gas ($\gtrsim 10^4$~K) is traced by 
\ion{C}{4}, \ion{Si}{4}, and \ion{N}{5}, but stellar winds superpose
broad absorption in these lines, making any interstellar component 
difficult to disentangle (Robert et al. 1993, Heckman et al. 1998).

In this paper, we measure interstellar gas kinematics toward star clusters
in a sample of nearby UV-selected galaxies.  In \S 2, we discuss the
{\em HST}/STIS observations, the data reduction, absorption line
diagnostics, and systemic velocities.  In \S 3 we discuss 
the ISM kinematics, including the kinematics of low-ionization
absorption lines in individual galaxies. In \S 4, we examine the
outflow trends across the sample, including the dependence of outflow
velocity on ionization energy, rotation speed, and star formation
rate.  In \S 5 we present a composite spectrum for the sample, examine
the high-ionization absorption lines, and compare the composite
spectrum to the high-redshift Lyman Break Galaxy composite
spectrum. The final section summarizes our results. We use
heliocentric velocities and H$_o$~=~75~\kms~Mpc$^{-1}$ throughout this
paper.    

%__________________________________________________________________________

% Section 2
\section{The Data}

% Section 2.1
%-----------------
\subsection{Observations, Data Reduction, \& Spectral Extraction\label{obsns}}

We have obtained Space Telescope Imaging Spectrograph (STIS)
long-slit far- and near-UV spectra for 17 nearby starburst
galaxies.  We target individual UV-bright nuclear or near-nuclear
(within 1~kpc of the galaxy center) star-forming clusters in the galaxies
to provide a strong background continuum source, against which to see
absorption. The galaxies were observed using the 52\arcsec $\times$
0.2\arcsec\, longslit and the G140L (FUV) and G230L (NUV) gratings,
projected onto the 25 arcsec$^2$ MAMA detector.  This configuration
provides continuous wavelength coverage from 1175~\AA~to 3100~\AA. The
STIS MAMA detectors have an average pixel scale of 0.584
(1.548)~\AA~pixel$^{-1}$ for the G140L (G230L) detector.  The spectral
resolution (FWHM), according to the STIS instrument handbook, is 8
pixels, or 0.2\arcsec, for a fully extended source. 

The galaxies in this sample were selected on the basis of high UV
luminosity; this is a tracer of star-formation.  The sample was
selected to include a range of galaxy morphologies and metal content
(7.7  $\lesssim$ log(O/H)+12 $\lesssim$ 9.2). See Table \ref{proptab}
for salient properties of the sample galaxies.  The young, hot stars
in the clusters provide a strong UV background source against which we
can see absorption by the ISM. Since these clusters are the most
UV luminous, almost all of them have ages of 3-5 Myr (Chandar et
al. 2004). The individual star clusters in these galaxies are the same
ones discussed by Chandar et al. (2004); they analyzed the absorption
present in the stellar photospheres and winds to characterize the
stellar population in each galaxy.  We concentrate on the interstellar
contribution to the spectra.  Our sample differs slightly from the
Chandar et al. (2004) sample. They include NGC~3049, NGC~5253 and
Tol~89, which we do not because the data were taken with a different
slit and are not part of this program.

We define dwarfs to be those galaxies with $M_B \ge -18$ (following
the convention of Thuan \& Martin 1981, Marlowe et al. 1997, and
numerous other authors), and separate our sample into two subsets in
which to study outflow kinematics: dwarfs and disks.  However, since
our sample is UV-selected, and thereby selected for the presence of
active star-forming regions, we expect to see mergers and interactions
in our sample, and the disk/dwarf classification scheme is less than
perfect. To this end, NGC~5102 is anomalous, since it is a lenticular
galaxy, classified as type SA0 pec (i.e., not a dwarf) but it has an
absolute blue magnitude of -17.11, which would place it in the dwarf
subsample.  Mkn~33 and Tol~1924-416 both have M$_B <$~-18, yet they
are classified in the literature as blue compact dwarfs.  Because of
this dichotomy in classification, these three galaxies are separated
from the disk and dwarf subsamples in this paper. 

The raw data files were retrieved from the {\em Hubble Space Telescope
(HST)} archive.  They were then co-added and processed through the
CALSTIS pipeline, which performs global detector linearity
corrections, dark subtraction, flat-fielding, wavelength calibration,
and conversion to absolute flux units.  We identify the same clusters
as Chandar et al. (2004), and extract the spectra using the X1D
package in IRAF\footnote{IRAF (Image   Reduction and Analysis
  Facility) is distributed by the National Optical Astronomy
  Observatories, which are operated by AURA, Inc., under cooperative
  agreement with the National Science Foundation.}.  This technique
extracts a spectrum and subtracts a background spectrum, usually
removing most of the geocoronal Ly$\alpha$ and \ion{O}{1}
$\lambda$1302 emission. The spectra were normalized by dividing each
G140L spectrum by the mode of the flux between 1250~\AA~and 1500~\AA.
The G230L spectra were normalized by dividing the spectra by a
third-order polynomial fit between 2000~\AA~and 3000~\AA.
Since we are interested in the kinematics of the interstellar
absorption lines, we retain the highest possible spectral resolution
and do not rebin the spectra.     

Multiple clusters were observed in every galaxy except NGC~5102 and
NGC~6764. If measured velocities for clusters in the same galaxy are
the same (i.e., a difference in velocity cannot be measured within 1$\sigma$), then the
individual cluster spectra are co-added to produce a single spectrum
with higher signal-to-noise.  This is the case in all galaxies with
multiple clusters extracted, except for NGC~3310.  When multiple
cluster spectra were averaged, the instrumental resolution for the
individual clusters were weighted by flux and averaged.
Table~\ref{kintab} provides the number of clusters averaged (Col. 2),
as well as the maximum offset of the clusters from the slit center in
either direction (Col. 3).   (The slit positions themselves are
presented in Table~2 of Chandar et al. 2004.)  Spatial variation of
the outflows along multiple sightlines is further discussed in
\S~\ref{space}.

% Section 2.2
%--------------
\subsection{Spectral Resolution\label{resn}}
%--------------
The spectral resolution for a point source observed with the
$0.2\arcsec$ STIS slit is $\sim$0.88~\AA~ at 1500~\AA; for a source
that fills the slit, the resolution is 4.67~\AA.  Most of the clusters
in our sample are neither point sources nor do they fill the slit.  We
measured the cluster surface brightness profiles from the
two-dimensional spectra by summing up 500 columns (12$\farcs$5) along
the slit spatially.  We can equate the resulting spatial scale in
pixels to an instrumental line width in Angstroms using the scale of
0.584~\AA~pixel$^{-1}$ in the G140L spectra.  Most clusters have an
instrumental FWHM between 1.5~\AA~and 3.5~\AA, with a  median size of
2.7~\AA~(or 607~\kms~at 1335~\AA).  These values (column~7 in
Table~\ref{kintab}) give a reasonable estimate for the resolution of
individual cluster spectra (i.e., the instrumental width).  

The instrumental width is subtracted in quadrature from
the measured width; line widths for the \ion{C}{2}~$\lambda$1335 line
are presented in column~8 of Table~\ref{kintab}. For the four galaxies
(NGC~1741, NGC~5996, NGC~7552, and Tol~1924-416) where we detect
foreground/Milky Way halo lines, the instrumental line width is
consistent with the widths in Table~\ref{kintab}. 
 
The wavelength calibration will be affected if the extracted cluster
does not fall in the middle of the slit (along the dispersion
axis). From the STIS handbook, typical wavelength calibrations are
accurate to $\sim$0.2 pixels, or $\sim$20~\kms~at 1500~\AA.  If the
cluster is shifted by 1/4 of the slit width, or 2 pixels, then the
wavelength calibration will be ``off'' by $\sim$200~\kms~at 1500~\AA.
This shift is large enough that it is easily discernible from
wavelength measurements of the stellar lines.  Comparison with stellar
lines in our spectra to literature values confirm that the spectra used in this work do not
suffer from any obvious zero point offset in the wavelength scale.

% Section 2.3
%--------------
\subsection{Systemic Velocities\label{systemics}}
%--------------
Once the spectra are extracted, the kinematics are examined.  In order
to measure the kinematics of the warm/cold gas in the ISM, we 
need to compare the systemic velocity of each object with the
velocity of the low-ionization lines.  The values of $v_{sys}$ for
each object (i.e., the barycentric velocity combined with the rotation
velocity at the cluster position) are given in Table \ref{kintab}.   
Ideally, the systemic velocity is derived from stellar lines in a
high-resolution spectrum. Keck\footnote{The HIRES data referenced
  herein were obtained at the W.M. Keck Observatory, which is operated
  as a scientific partnership among the California Institute of
  Technology, the University of California and the National
  Aeronautics and Space Administration. The Observatory was made
  possible by the generous financial support of the W.M. Keck
  Foundation.}/HIRES (Vogt et al. 1994) optical spectra of He~2-10,
NGC~4214, and NGC~4449 provide a systemic velocity from the
\ion{Mg}{1} b-band absorption triplet, which is present in the K-type
giants and supergiants in the starburst region.  When high-resolution
spectra of the star clusters are not available, a systemic velocity is
taken from the literature/NED. Preference is given to CO and \ion{H}{1}
velocities over emission-line velocities since the CO and \ion{H}{1}
likely trace the gas in the colder molecular clouds in which the
starburst occurs, while the emission lines could be coupled to an
outflow or other non-stellar process.  When the value of $v_{sys}$ in
the literature is not in agreement with the stellar lines seen in the UV
spectrum, the stellar velocity is used; this is the case for NGC~3125,
NGC~7552, and Tol~1924-416.  (Stellar lines used include, but are not
limited to, \ion{C}{3}~$\lambda$1427, \ion{Fe}{5}~$\lambda$1431,
\ion{S}{5}~$\lambda$1502, and \ion{He}{2}~$\lambda$1640.)  
%The value
%of $v_{sys}$ used for all other galaxies is consistent with the values
%given in Table~\ref{kintab}. 

% Section 2.4
%--------------
\subsection{Spectral Line Identification\label{lineid}}
%--------------

Table \ref{linetable} presents the wavelengths and ionization energies
of lines measured in the spectra.  Severely blended lines are noted.
In the G140L/FUV grating, the best lines to use to characterize the
purely interstellar gas (i.e., the least blended and strongest lines)
are \ion{Si}{2} (1260~\AA), \ion{C}{2} (1335~\AA), \ion{Si}{2}
(1527~\AA), and \ion{Al}{2} (1671~\AA). Although Ly~$\alpha$
(1216~\AA) is an important diagnostic line, our local spectra are
contaminated by strong airglow emission from the earth's atmosphere,
which makes it very difficult to disentangle/analyze the underlying Ly
alpha emission and absorption from the starburst galaxy itself.  The
\ion{O}{1} (1302~\AA) and \ion{Si}{2} (1304~\AA) lines are blended,
and the line combination can be affected by geocoronal \ion{O}{1}
emission if the background/sky subtraction is imperfect; therefore we
exercise caution when fitting this line.  We follow the convention of
Shapley et al. 2003, and use an average wavelength for
\ion{O}{1}/\ion{Si}{2} of 1303.27~\AA\, for this line.  The G230L/NUV
spectrum contains fewer lines, which tend to be weak. \ion{Mg}{1}
$\lambda$2853, is a blend of stellar and interstellar absorption,
though it is only detectable in six galaxies. The \ion{Mg}{2} doublet
near 2800~\AA\, is largely interstellar, and is detected in nearly all
the galaxies though it is often difficult to deblend the doublet
lines.  A multitude of blended \ion{Fe}{2} lines are present from
2344~\AA\,to 2600~\AA.

The lines are all fit with Gaussian profiles using the {\em splot}
task in IRAF.  This task does not allow us to constrain the line fits
(e.g., all \ion{Si}{2} lines should have the same velocity and width),
however it is a reliable method for fitting spectra at this
resolution.  To check the accuracy of our fitting, three galaxy
spectra were also modeled with the SPECFIT program (Kriss 1994), which
does constrain line parameters, and also produced an output model
spectrum.  The results differed by $\lesssim$20\%.  The
lines were therefore fit with {\em splot};  all low-ionization
interstellar line velocities were averaged to give an
overall low-ionization interstellar outflow velocity.  

%__________________________________________________________________________

% Section 3
\section{Results\label{results}}

The results are presented in Table \ref{kintab}.  The resulting
difference between the systemic/stellar velocity and the velocity of
the low-ionization lines measures the outflow velocity of the cool gas
entrained in the galactic wind (if any).   Twelve of the 14 
galaxies with detectable interstellar absorption lines show blueshifts,
and two do not.  The spectra of NGC~4449, NGC~4861, and VII Zw
403 are too noisy to detect any absorption.  While NGC~4214 shows an 
outflow based on other data (e.g., Schwartz \& Martin 2004), our data do 
not resolve this.  The other four galaxies (NGC~1741, NGC~3125,
NGC~6764, and Tol~1924-416) are consistent with zero velocity (within 
1-$\sigma$).

%Of the galaxies with
%outflows seen, only two of the galaxies -- NGC~4214 and Tol1924-416 --
%could possibly have infalling gas (i.e., redshifted absorption) within
%the estimated uncertainty.  Three galaxies in the sample have
%essentially no detectable outflow; they are NGC~1741 ($\Delta
%v_{outflow} = 16\pm$76~\kms), NGC~3125 ($\Delta v_{outflow} =
%8\pm$73~\kms), and NGC~6764 ($\Delta v_{outflow} = -9\pm$50~\kms).  

%--------------
% Section 3.1
%--------------
\subsection{Doppler Shifts\label{dopps}}
%--------------

A low-velocity Doppler shift is difficult to detect, due to the
blending of lines and the modest resolution of the spectra. The accuracy
with which the absorption lines can be fitted depends on the noise in
the spectrum; on average, we estimate an error of $\sim$10\% of the
spectral resolution, or about 80~\kms~in fitting the kinematic center of a
spectral line.  
Figure \ref{snrerror} shows the relation between signal-to-noise
ratio per Angstrom, (S/N)$_{\AA}$, and the fitting error, $\delta v$:
\begin{equation}
\delta v(\mbox{km s}^{-1}) = -1.8\times(S/N)_{\AA} + 104\mbox{ km s}^{-1}
\end{equation}
We find that for a typical S/N of 13, the error in fitting is
81~\kms, which agrees with our estimate.  For a cleaner spectrum with
S/N of 30 (per pixel), the error is 50~\kms, and for a noisier
spectrum with S/N of 8, the error is 90~\kms.    

The velocities of all detectable interstellar lines are measured and
the average outflow velocity $\Delta v_{outflow} = v_{sys} - v_{line}$
is presented in Table \ref{kintab}.  \ion{C}{2} $\lambda$1335,
\ion{Si}{2} $\lambda$1260, and \ion{Al}{2} $\lambda$1671 are the only
three unblended lines which are present in every spectrum.  \ion{Si}{2}
$\lambda$1527, while blended with the high-ionization \ion{C}{4}
doublet, is detected and measured in all spectra.  The \ion{C}{2} and
\ion{Si}{2} spectra are presented in Figure \ref{ciifig}.  

We detect outflowing gas in eight of the 17 galaxies.  The detection
threshold of 1-$\sigma$ may seem rather liberal; however, a threshold
of 2-$\sigma$ would decrease the detection level from eight galaxies
to seven.  The galaxy which does not fall into the 2-$\sigma$ level is
NGC~5102.    

A velocity shift of about 80~\kms~will be detected,
particularly if it is present in multiple interstellar absorption
lines. If there are multiple lines of a species (e.g., \ion{Si}{2}
$\lambda$1260, $\lambda$1527), we know those lines should have the same
velocity and width.  Moreover, if ions have similar ionization energies (see
Table~\ref{linetable}), the corresponding absorption lines should
arise from regions of gas at similar temperatures, and therefore
should be spatially coincident. The outflow velocities (and errors
based on the signal-to-noise in the spectrum) are presented in
column 6 of Table \ref{kintab}.  The highest-velocity outflows are clearly 
blueshifted from the systemic velocity. Projection effects due to a 
non-zero inclination angle (column~6 of Table~\ref{proptab}) are not
considered because we do not know the opening angle of the outflow.

%--------------
% Section 3.2
%--------------
\subsection{Spatial Variation of Outflows\label{space}}
%--------------

The spatial variation in outflow speed along the slit is interesting
since some starbursts show a more obviously global outflow. For example, many
ultraluminous infrared galaxies show a coherent outflow over kpc scales
(Martin 2005).  Systems without an outflow observed near
one cluster might show outflows along other sightlines.   

At high resolution, two distinct clusters in the dwarf
galaxy NGC~4214 (NGC~4214-1 and -2, separated by $\sim500$~pc) show
different kinematics measured by optical Na D absorption (Schwartz \&
Martin 2004).  NGC~4214-2 shows a blueshift of just -23~\kms, whereas
NGC~4214-1 shows no shift at all, demonstrating the small (yet
detectable and significant) differences in outflows between star
forming clusters in the same galaxy.  While this does not show that
the outflows are tracking local chimneys from the disk, similar
outflow velocities from spatially separated clusters can demonstrate
the scale of spatial coherence in an outflow. Systems without an
outflow observed near one cluster might show slightly different
outflow kinematics along other sightlines, although at this resolution
we cannot distinguish $\sim20$~\kms differences.

NGC~4214-2 was not observed with STIS.  When NGC~4214-1 is fit with
SPECFIT, the interstellar lines (\ion{C}{2}, \ion{Al}{2}, \ion{Si}{2})
converge to a fit where they are blueshifted by -32~\kms~$\pm$
49~\kms. Since there is no outflow detected in the optical spectrum of
NGC~4214-1, it is not surprising that the UV velocity measurement is
consistent with no outflow.   While by no means conclusive, this
marginally demonstrates that outflows can be localized, although
within the errors the UV data of NGC~4214-1 does not rule out a slow
outflow (as in NGC~4214-2).  

Multiple sightlines could be investigated in He~2-10.  It has two main
starburst regions, A and B, which are separated by 350~pc, and shows a
large outflow velocity (M\'endez et al. 1999). Johnson et al. (2000)
found similar outflow velocities from regions A and B. Our spectra
from region A, the central starburst which contains five bright
clusters, are shown in Figure \ref{he210}.   The regions all show
similar kinematics, so we combine the spectra to increase the
signal-to-noise.  This is the case with every galaxy showing multiple
clusters with the exception of NGC~3310.

NGC~3310 is the only galaxy which exhibits significant ($\gtrsim
1\sigma$) kinematic differences in clusters -- regions A, B, and C
have different outflow velocities, and are therefore examined
separately.  The kinematics of these regions are presented in
Table~\ref{kintab}.   One of the larger and more complex
galaxies we observe is NGC~3310. The STIS slit is positioned to
observe three large clusters near the center of the galaxy -- see
Figure \ref{3310map}.  The de-redshifted spectra of regions
NGC~3310-A, -B, and -C are shown in Figure \ref{diskspec}. The spectra
of the interstellar lines (dashed/red) make it clear that not only is
there a significant outflow of interstellar gas from the star-forming
regions, but this outflow changes across the slit.   Region A is
$\sim$850 pc from region C, which is $\sim$100 pc from region B.  The
\ion{H}{1} velocities are the same in Regions B and C, whereas Region
A is more redshifted by 30~\kms (Mulder 1995); this is too small a
velocity difference to be detected in these spectra, as discussed
above. However, there are large ($\ga$100~\kms) differences in the
outflow speeds from these three regions, with Region A (C) being the
slowest (fastest) outflow; the slowest outflow is still significantly
larger ($\Delta v$ = -339~\kms) than any other Doppler shift in this
sample. This suggests we are observing either local outflows from
different regions of the disk (rather than some global outflow), or
local variations {\em within} a global outflow.  The scale over which
kinematics differ is surprisingly small -- regions NGC~3310-B and -C
are separated by only $\sim$100~pc.

%--------------
% Section 3.3
%--------------
\subsection{Comparison to Previous Results\label{prev}}
%--------------

Previous measurements of outflow velocities from UV absorption lines
are available for only three galaxies in our sample (e.g., He~2-10:
Chandar et al. 2003, Johnson et al. 2000; Mkn~33 \& Mkn~36: Kunth et
al. 1998).  He~2-10 is discussed below. The outflow velocity obtained
for Mkn~33 agrees well with the measurement of Kunth et al. (1998).
Mkn~36, however, does not agree: Kunth et al. (1998) find a redshift
of +40~\kms~whereas we find a relatively fast outflow.  This may be
due to the availability of many more lines in our spectrum; Kunth et
al. (1998) were only able to measure two absorption lines, \ion{O}{1}
$\lambda$1302 and \ion{Si}{2} $\lambda$1304. 

Larger galaxies with a more powerful starburst region tend to show
a larger and therefore detectable Doppler shift (see \S4.1 and \S4.3).  For example, NGC
7552 is a type SBab spiral galaxy.  The STIS slit was placed across the
nucleus. Figure \ref{7552fig} shows the absorption spectrum of this
galaxy; a blueshift of $-316$~\kms~$\pm$ 50~\kms~is seen in the three
unblended absorption lines. This is marginally consistent with the
observations of Heckman et al. (2000), where the cold neutral medium
is observed via Na D absorption, and is measured to have a blueshift
of $-216$~\kms~$\pm$~20~\kms.  

The five separate clusters in region A of He~2-10 have the same outflow
velocities to within $\lesssim 30$~\kms, (the fainter clusters have lower
S/N and hence larger uncertainties).  The low-ionization lines are fit
with the SPECFIT program; this allows us to constrain the model fit,
forcing lines of the same species to have the same velocity and
width. Using the systemic velocity for this region from Mg-b
absorption in high-resolution optical spectra (Schwartz \& Martin
2005, in prep.), we fix the systemic velocity to be at the same
redshift as the stellar (Mg-b) optical absorption.  We measure an average UV
low-ionization outflow speed of $-170$~\kms~$\pm$ 8~\kms~from the
central starburst (commonly called Region A). This velocity is lower
than the $-360$~\kms~outflow (from \ion{C}{2} and \ion{Si}{2})
reported by Johnson et al. (2000). Part of the discrepancy, about
30~\kms, is due to our systemic velocities; but most of the difference
is likely due  to the uncertain position of the clusters in their
larger aperture. In our recently obtained optical spectrum of He~2-10,
the interstellar \ion{Na}{1} absorption shows an outflow with four
separate wind components (Schwartz \& Martin 2006, in prep.). The
highest velocity component is at $\Delta v_{Na} = -128$~\kms, which is
consistent with the low-ionization UV absorption measurements
presented here.  

%--------------
% Section 3.4
%--------------
\subsection{Line Widths\label{widths}}
%--------------

It is difficult to analyze the (Gaussian) full-width at half-maximum
(FWHM) the same way we analyze the line velocities.  When centering a
line, there is usually an obvious absorption minimum, as seen in the
\ion{C}{2} spectra presented in this paper. Therefore, measuring the
kinematic center of a line is generally straightforward.  However, it
is more complicated to accurately measure the FWHM since the multiplets
are often blended with other species arising from different ionization
energies and therefore possibly different gas phases.  For example,
\ion{Si}{2} $\lambda$1527 (ionization energy 8.15 eV) is blended
with the \ion{C}{4} $\lambda\lambda$1548, 1551 lines (ionization
energy 47.89 eV), which often shows an extremely broad P Cygni
profile.  For this reason, we use the FWHM of the unblended \ion{C}{2}
$\lambda$1335 line (ionization energy 11.26 eV) to characterize the
FWHM of the gas in the cold/warm phase of the ISM (see column~8 of
Table~\ref{kintab}).  All references to FWHM herein imply a
deconvolved, deredshifted width.

The deconvolved line widths (measured at the Gaussian FWHM) measured
range from few$\times$100 \kms~to $>1000$~\kms; these are broader than
the Na D line widths seen in dwarfs (Schwartz \& Martin 2004), LIRGs
(Heckman et al. 2000) or in ULIRGs (Rupke et al. 2002, Martin 2005).
The thermal line width is a few~\kms~for a gas at 10$^4$ K.  
%At high resolution the broad absorption profiles of cold gas in the
%nearby starbursts NGC 1614 and M82 break up into multiple smaller
%components, each still broader than the thermal line width (Heckman
%et al. 2000, Schwartz \& Martin 2006).  
The lines we observe are significantly
broader than the instrumental resolution, although we do not know this
resolution precisely.  The absorption lines in the STIS spectra are
not resolved. 

The estimated maximum/terminal velocity we see scales roughly as
$\Delta v_{outflow} + FWHM/2$; see Figure~\ref{fwhmfig}.  (The exact
fit is $0.62\pm0.17\times$ FWHM(\ion{C}{2}).  The profile shapes are
described by a Gaussian profile.  Generally, the lines get broader as
the outflow velocity increases in magnitude. This is consistent with
the physical picture of gas clouds being injected into a flow at the
systemic velocity and being accelerated, as seen by Heckman et
al. (2000).  Since the instrumental FWHM comes from the spatial size
of the cluster (see \S~\ref{resn}), even with the largest instrumental
widths we still see a non-zero FWHM in the \ion{C}{2} line. This 
suggests that perhaps we are seeing a conglomeration of filaments and
shells of interstellar gas at a variety of velocities, as seen in
simulations (Fujita et al. 2005, in prep.), rather than a single cloud
or simple shell.   

%__________________________________________________________________________

%--------------
% Section 4
%--------------
\section{Kinematics of the Interstellar Medium -- Analysis\label{kmix}}
%--------------

In Figure~\ref{params}, we plot the outflow velocity $\Delta v$
against the cluster age, UV slope, and metallicity.  There are no obvious
correlations among these parameters seen in the data, whether or not
the samples are divided into sub-samples by galaxy type (i.e. dwarf or
disk).   Therefore, outflow velocities are likely more dependent on
the potential well of the galaxy and the star formation rate than the
properties of the stars in the starburst.  Surface brightness, star
formation rate, and/or cluster mass may be the best parameters to
examine, since the outflow seems more dependent on the galactic-scale
environment than the stellar population.  

%--------------
% Section 4.1
%--------------
\subsection{Rotation Speed vs. Outflow Velocity\label{rots}}%{\bf [HELP!]}}
%--------------
An important query is whether or not the outflowing interstellar
matter is moving fast enough to escape the gravitational potential
well of the host galaxy.  We can parameterize the dynamical mass and
therefore the gravitational potential of the galaxies with the
galactic rotation speed.  The escape velocity ranges from
$\sqrt{2}v_{rot}$ (minimum) up to $\sim 3v_{rot}$ (for gas extending
$\sim 3$~kpc from the nucleus in an isothermal halo out to 100~kpc);
a typical dwarf halo escape velocity is 100~\kms~ (Martin 1998).  The
resulting terminal velocities for the interstellar lines in our
spectra are consistent with outflow being accelerated up to the escape
velocity, but generally with no material exceeding $v_{esc}$. We plot the
outflow velocity as a function of rotation speed in Figure~\ref{vtvr}.
The galaxies are separated into subsamples as described in
\S~\ref{obsns}; the bright nearby dwarf starburst galaxy NGC~1705
(V\'azquez et al. 2004) is plotted as well.  

Using the rotation speed as a proxy for halo mass, the overall trend
suggests that more massive halos produce faster outflows.  A
least-squares fit gives a slope of $1.2\pm 0.2$.  The terminal
velocity ($v_{term}\sim \Delta v + 0.5\times$FWHM) may exceed the
escape velocity significantly in two dwarf galaxies in particular --
Mkn 36 and Mkn 209.  This tentatively suggests that entrained outflow
material from the ISM is likely to escape from these smaller galaxies.
This result agrees with the debated theoretical picture that dwarf
galaxies may be more susceptible to mass loss via superwinds, simply
because they have a smaller potential well (e.g., Larson 1974; Dekel
\& Silk 1986; Martin 1998; Ott, Walter \& Brinks 2005).  

As an example, the rotation speed of NGC~7552 from \ion{H}{1}
measurements is 230~\kms; we estimate the terminal velocity of the gas
in the outflow is $\Delta v + 0.5\times$FWHM $\sim$ 700~\kms.  This
indicates that using a reasonable estimate for the escape velocity of
$v_{esc} \sim 3v_{rot}$ (for an isothermal halo at 100~kpc and gas
extending $\sim3$~kpc; Binney \& Tremaine 1987), the escape velocity
is $v_{esc}\sim$700~\kms, and it is not likely that the gas is moving
fast enough to reach the escape velocity for the galactic potential.

%--------------
% Section 4.2
%--------------
\subsection{Kinematics and Ionization Energy}
%--------------

An important objective of this work is to check the consistency of
outflow speeds measured from different lines.  We expect ions with
higher ionization energies to trace higher-temperature gas, so hotter
gas exhibits a faster outflow velocity (V\`azquez et al. 2004).  This
relationship agrees with the scenario of hotter, less dense gas
breaking free from a starburst region and expanding outward at a
higher rate than the colder, more dense gas from the ISM that is
loaded into the flow (e.g., Schwartz \& Martin 2004 and references
therein).   

High-resolution results (V\'azquez et al. 2004) show a direct
relationship between ionization energy and outflow velocity. In
Figure~\ref{cvsal}, we show the outflow velocities for \ion{Al}{2} and
\ion{Si}{2} versus \ion{C}{2} velocity.    Weighted least-squares fits
to the \ion{Al}{2} vs. \ion{C}{2} data give a slope of $0.61 \pm 0.13$.
The correlation is noisy, but the \ion{Al}{2}-absorbing gas seems to
be moving slower than the \ion{C}{2}-absorbing gas.  The
\ion{Si}{2}-absorbing gas is less conclusive, with a slope of
$0.77 \pm 0.15$. However, the overall trend of $\Delta v_{CII} > \Delta
v_{SiII} > \Delta v_{AlII}$ is consistent with IE(\ion{C}{2}) $>$
IE(\ion{Si}{2}) $>$ IE(\ion{Al}{2}) (where IE is the energy required
to remove one electron and create the ion in question, as listed in
Table~\ref{linetable}.    

Optical interstellar \ion{Na}{1} absorption lines have been detected
for three galaxies examined - NGC~4214, NGC~7552, and He~2-10, as
discussed in \S~\ref{prev}.  The
low-ionization UV lines for NGC~4214 (Fig.~\ref{n4214fig}; Schwartz \&
Martin 2004) and NGC~7552 (Fig.~\ref{7552fig}; Heckman et al. 2000)
yield velocities slightly faster than the \ion{Na}{1} measurements,
although the optical spectra constrain the outflow speed better.  The
UV lines provide good constraints for He~2-10, where the average of
several low-ionization lines is roughly consistent that of the highest
velocity component seen in the \ion{Na}{1} spectrum.  The slightly
faster speeds seen from the low-ionization UV lines (as compared to
\ion{Na}{1}) are consistent with the average I.E. of the UV lines
being greater than I.E. of \ion{Na}{1}.  These results show that
$\Delta v_{NaI} < \Delta v_{CII}$ in galaxies where both lines are
measured; this is in agreement with the absorption from different ions
occurring in physically distinct regions of the outflow.

%--------------
% Section 4.3
%--------------
\subsection{Star Formation Rate vs. Outflow Velocity\label{sfr}}
%--------------

Another important parameter to investigate in relation to outflow
velocity is the star formation rate (SFR).  In Figure~\ref{sfrfig}, we plot
outflow velocity versus SFR, as measured by \ha luminosity (column 11
of Table~\ref{proptab}), using the following equation from Kennicutt
(1989): 
\begin{equation}
\mbox{SFR~(M}_\odot \mbox{ yr}^{-1}) = \frac{L_{H\alpha}}{1.26\times 10^{41}}.
\end{equation}
Although these \ha measurements were not corrected for internal extinction,
the implied SFRs were found to be roughly consistent with those derived from
the measured UV luminosities.
%\footnote{
%  The {\em IUE} UV spectra from Kinney et al. (1993) were used to measure
%  the  UV flux for most of the galaxies. The data for 
%  He~2-10 and NGC~1741 were taken from Johnson et al. (2000) and Conti et al. 
%  (1996), respectively.  To account for the UV spectral slope, $\beta$
%  (column 9 of Table~\ref{proptab}), we followed Bell \& Kennicutt
%  (2001).} 
In contrast, the SFRs estimated from the FIR {\em IRAS} fluxes in the
60- and 100-micron bands were generally significantly lower than the
UV-\ha-derived SFRs.  Hence we adopted the \ha estimates of the global
SFR.    

A general trend is apparent: overall, galaxies with a higher SFR
exhibit faster outflow speeds. A simple least-squares fit gives a
slope of $0.51\pm 0.59$, This is consistent with the results for
\ion{Na}{1} velocity versus SFR presented in Martin (2005), and shows
that UV absorption lines and optical resonance lines yield consistent 
results. However, defining this relation empirically would benefit
from more and higher resolution spectra.  Also, since the SFRs are
quoted for the entire galaxy here, more insight could be gained from
a comparison to surface photometry of the individual star-forming
regions. The total SFR is much greater than that of the cluster(s)
observed in the largest  galaxies such as NGC~3310. 

The galaxies in this sample are intermediate in luminosity to the
dwarfs in the Schwartz \& Martin (2004) sample and the LIRGs in the
Heckman et al. (2000) and Rupke et al. (2004) samples.  The $v - SFR$
relation of Martin 2005 predicts that velocities will level off around
$10^1 - 10^2$~M$_\odot$~yr$^{-1}$, which is around the SFR of LIRGs.
Therefore, the fitted velocities should be near or slightly less than
the outflow velocities in the LIRGs; in general, they are.

%__________________________________________________________________________

%--------------
% Section 5
%--------------
\section{A Local Star-forming Galaxy Composite Spectrum\label{composite}}
%--------------

To improve S/N, and study weaker lines as well as the systematic
properties of the local star-forming regions, we created a composite
spectrum of all the individual spectra.  To produce our composite
spectrum, we first shift all individual G140L spectra into the rest
frame using the systemic velocities in Table \ref{kintab}.  We then
normalize each of the 16 individual spectra to the mode between
1250~\AA~and 1500~\AA. The spectra are averaged and rebinned to a
dispersion of 1~\AA.  The composite local starburst spectrum is
presented, along with the composite LBG spectrum from Shapley et
al. (2003), in Figure~\ref{lbgfig}.

%--------------
% Section 5.1
%--------------
\subsection{Kinematics of the Local Composite Spectrum}
%--------------

% CHARACTERISTICS OF COMPOSITE SPECTRUM.
% Low-ionization lines blueshifted.  
% High-ionization lines -- NV blended/not there.  P Cygni profiles
%                          strong, can't see obvious IS component.
%                          Compare SiIV and CIV to SB99 model, don't
%                          need to add IS component to accurately
%                          describe lines; could be small amt of IS
%                          but can't say definitely... Vazquez+ *do*
%                          see IS CIV/SiIV in high-res'n spectrum of
%                          N1705.
% 

The low-ionization absorption lines are blueshifted in the local
composite spectrum.  While the lines are still blended due to stellar
emission/absorption, there is little absorption redward of the
systemic velocity, while blueward there is strong absorption.
We used {\em splot} to fit Gaussian profiles to the low-ionization
interstellar absorption lines (\ion{Si}{2}, \ion{C}{2}, \ion{Al}{2},
\ion{O}{1}/\ion{Si}{2}) and found an average low-ionization outflow
speed of $\Delta v_{outflow} = -142$~\kms~$\pm$~80~\kms.  

In the composite spectrum, the higher S/N allows us to look for
absorption from more highly ionized ions. There are three
high-ionization lines: \ion{Si}{4}, \ion{C}{4}, and
\ion{N}{5}. \ion{N}{5} is not present in the majority of the spectra,
and when detected it is blended with Lyman~$\alpha$~absorption and
corrupted by the residuals from sky subtraction of Lyman~$\alpha$
emission.  
%When compared to a Starburst99 (Leitherer et al. 1999)
%model spectrum (5 Myr, solar metallicity, instantaneous burst, 1-100
%M$_\odot$), 
There is no distinct interstellar absorption component in
the \ion{Si}{4} or \ion{C}{4} lines.  The local composite
\ion{C}{4} and \ion{Si}{4} are accurately described by a Starburst99
model spectrum (1-100 M$_\odot$, 0.25Z$_\odot$ metallicity,
instantaneous burst at 5 Myr; Leitherer 1999); we cannot confidently
discern a high-ionization interstellar component in the composite
spectrum in any high-ionization line.  
%The P Cygni profiles of the stellar lines adequately describe the
%high-ionization lines in the composite spectrum at this spectral
%resolution. 
However, note that there is a detection of blueshifted interstellar
\ion{N}{5}, \ion{Si}{4} and \ion{C}{4} in the high-resolution
spectrum of NGC~1705 (V\'azquez et al. 2004); better resolution
spectra of the clusters in this paper would likely yield a detection
of some interstellar gas in these higher ions.  

%__________________________________________________________________________

%--------------
% Section 5.2
%--------------
\subsection{Comparison with the High-$z$ Composite Spectrum\label{compz}}
%--------------
% COMPARE TO LBG SPECTRUM
% Comparing our composite to theirs.  The two samples are related b/c
%     both SF galaxy samples, but they have 811 to our 17, plus the
%     luminosities are different.  
% Initially surprisingly similar.  Continuum slopes are practically
%     the same, low-ionization lines seem to have ~same outflow
%     velocities. Also lines that initially look like noise in either
%     spectrum seem to line up, therefore might be weak metal lines.

We compare our composite spectrum to the composite spectrum of the
LBGs presented by Shapley et al. (2003).  The samples differ in the
number of galaxies included in the composite; Shapley et al. average
811 galaxies, whereas we only include 17 galaxies.
Figure~\ref{luvhist} shows a histogram of the UV luminosities of the
two samples; the local sample (average log \luv = $38.82$
erg~s$^{-1}$) does not overlap at all with the LBG sample
(average log \luv = $40.99$ erg~s$^{-1}$).  Moreover, there
is an ``aperture bias'' between the local and high-$z$ samples. In the
local spectrum we disproportionately weight the light from the
clusters (which represent the youngest stellar population), whereas
the high-$z$ composite includes spectral features from both
star-forming clusters and a diffuse ``field'' component.  Chandar et
al. (2004, 2005) discuss in detail the contribution of the more
diffuse ``field'' population to composite spectra.  

Given the differences in the two samples, the similarities between the
local composite spectrum and the LBG composite spectrum are
striking. First, the slopes in the low- and high-$z$ composite spectra
are remarkably similar: roughly -1.35 and -1.5, respectively (slope
$\beta$ between 1240~\AA~and 1600~\AA~for $F_\lambda \propto
\lambda^\beta$). Using Starburst99 data (Leitherer et al. 1999), we
can look at how the UV slope in a starburst changes with time. Both
instantaneous and continuous bursts show a constant slope as a
function of time out to $\sim$10$^{7.2}$ years, so the similar slopes
are not hard to produce. The low-ionization interstellar absorption
lines also show similar average outflow velocities: $\Delta
v_{outflow}$~=~ $-142$~\kms~$\pm$~80~\kms~in the local sample, and
$\Delta v_{outflow}$~= ~$-150$~\kms~$\pm$~60~\kms~in the LBGs.

% HOWEVER, our velocity *distribution* is different -- our spectra
%     more weighted towards brightest/most luminous/highest SNR
%     galaxies which have biggest blueshifts, so dwarfs don't count as
%     much in the average.  So while the velocities are similar, it's
%     galaxies most like LBG luminosities who are most important in
%     our average.

The similarity in low-ionization outflow velocity is quite surprising
since all of the galaxies that comprise the high-redshift composite
spectrum have higher UV luminosity than even the most luminous member
of the sample presented here. However, the velocity in the local
composite is weighted towards the higher signal-to-noise spectra,
i.e. the more luminous clusters, which have intrinsically higher
outflow velocities.  Therefore the dwarfs, with their overall lower
outflow velocities, have less of an overall effect on the composite
than the brighter galaxies, whose higher luminosity and higher outflow
speeds are closer to that of the LBGs.  While the composite outflow
velocities may be similar in low-ionization lines between the two
samples, the velocity {\em distributions} are different.  Moreover, the
SFRs for the two samples are both at or above than the ``turnover'' in the
$v - SFR$ relation of Martin (2005).  Thus, an increase in SFR above
this level ($\sim 10^1 - 10^2$~M$_\odot$~yr$^{-1}$) will not result in an
increase in the outflow velocity.  

% Also, closer inspection shows our lines are broader than LBGs.  CIV
%     is broader because we don't include the field component, whereas
%     LBG samples whole galaxy, not just bright clusters.  Other lines
%     are broader because of a combination of (a) difference in
%     spectral resolution and (b) broader range across luminosity &
%     therefore velocities, including zero velocity. 

Further examination of the composite spectra reveals that the local
lines are somewhat broader than the high-$z$ lines.  This can largely
be explained by a difference in spectral resolution (local composite
resolution $\sim$800~\kms, $z\sim3$ composite resolution
$\sim$600~\kms).  The lines in the local composite are also sampling a
broader range in luminosity, including both galaxies with low outflow
velocity and those with large outflow.  The combination of these two 
effects produces broader lines in the local composite spectrum.

In particular, the \ion{C}{4} line in the local composite spectrum is
broader than the absorption in the LBG composite spectrum, even after
accounting for the spectral resolution.  We interpret this chiefly as an
effect of age and aperture (since we are looking at younger stellar
populations with STIS, as described above).  Both the local and high-$z$
composite spectra show a more pronounced P Cygni-like emission
component in \ion{C}{4} than in \ion{Si}{4}.  
%Shapley et al. (2003) marginally detect
%interstellar \ion{Si}{4}, although origin of the \ion{C}{4} component
%is unresolved (Chandar 2003).  
It is hard to reproduce the LBG \ion{C}{4} shape from Starburst99
models or local starburst galaxy components, but either the ``blue
edge'' or the emission portion can be fit with these stellar models
(Chandar et al. 2003).  Hence, higher resolution observations are also
required to definitively detect the high redshift outflows in high
ionization lines.  For example, the high resolution spectrum of the
lensed $z=2.7$ Lyman break galaxy MS~1512-cB58 (Pettini et al. 2002)
shows blueshifted interstellar \ion{N}{5}, \ion{Si}{4} and \ion{C}{4}
lines. 

As another interesting similarity between the local and high-$z$
composites, some weak lines which appear to be noise in one of the
spectra also are present in the other.  The nature of these lines
(e.g. the small lines at 1467~\AA, 1579~\AA, or 1689~\AA) is unknown,
although they may be weak metal lines.   Additionally, although the
resolution of the composite spectra is too low to study column
densities or covering factors via specific lines, the high resolution
spectra of the local dwarf starburst NGC~1705 (V\'azquez et al. 2004)
and the Lyman break MS~1512-cB58 show similarly saturated absorption
in the rest-frame UV.  

%LBG observations include light from both star clusters as well as the
%older, more metal-rich diffuse/field component, whereas our
%observations specifically extract only the light from the
%clusters. Therefore, we sample an overall younger population with
%lower metallicity than the LBG observations. Chandar et al. (2003;
%2005) discuss this in greater detail.     
%The composite LBG spectrum of Shapley et al. (2003) shows
%significantly  blueshifted high-ionization gas at redshifts $z\sim$3
%which they claim to be interstellar.    
%__________________________________________________________________________

%--------------
% Section 6
%--------------
\section{Conclusions\label{concs}}
%--------------

We have detected many outflows of cool/warm interstellar gas in {\em
HST}/STIS  ultraviolet spectra of star-forming regions in nearby
UV-selected galaxies.  While the spectral resolution of the G140L and
G230L gratings is modest,  eight of 17 galaxies 
%(i.e. those spectra with adequate S/N ratio) 
show unambiguous blueshifted absorption in low-ionization interstellar
lines at velocities from -100~\kms~to -520~\kms. Somewhat lower
velocities are measured for three of the dwarf galaxies, but higher
resolution spectra are needed to confirm these estimates as typical
measurement uncertainties are $\pm 80$~\kms.  More massive galaxies
with higher star formation rates generally host the higher velocity
outflows.  No redshifted interstellar absorption lines are found,
which indicates a lack of inflows.
%significantly contrains the properties of gaseous inflow models.
That the lines are relatively broad (average \ion{C}{2} line FWHM
$\sim 800$ \kms) is likely indicative of sightlines probing a
multitude of filaments and shells as seen in numerical simulations,
rather than a single shell or bubble.   

In practice, the velocities presented in this paper were heavily
weighted by the strong, unblended lines from \ion{C}{2}, \ion{Si}{2},
and \ion{Al}{2}. Since C and Si have ionization energies between those
of Na and H, it is possible that (like Na~I) \ion{C}{2} and
\ion{Si}{2} are tracing gas where a significant fraction of the H gas
is neutral.  We looked for velocity differences among the UV lines
with mixed results.  The \ion{Al}{2} line tends to present lower
velocities than the \ion{C}{2} line. Since the ionization energy of Al
is less than that of C, this result supports previous suggestions that
higher ionization gas moves faster than less ionized material
(V\'azquez et al. 2004).  The ionization energy of silicon lies
between that of aluminum and carbon, but the \ion{Si}{2} line
measurements show higher  velocities than the \ion{C}{2} line towards
galaxies where \ion{Al}{2}  presents lower outflow
velocities. Unfortunately, the high ionization lines from \ion{C}{4},
\ion{Si}{4}, and  \ion{N}{5} are too blended with stellar lines or
airglow to reliably measure an interstellar component in individual
spectra (but see the composite spectrum below). We conclude that the
velocities will need to be measured more precisely before any
definitive statements about velocity and ionization state can be made.   

To quantify spatial variations in the outflows, we measured line
velocities toward all clusters intersected by the STIS longslit. For
example, in NGC~4214 the \ion{Na}{1} spectra clearly show outflow
toward cluster \#2 but none toward cluster \#1 only 500 pc away (Schwartz \& Martin 2004). We
found substantial differences in outflow velocities toward 3 clusters
in an outer spiral arm of NGC~3310 on scales of 100 pc and 900 pc. In
contrast, large velocities are measured all across the dwarf starburst
He~2-10. Indeed, in all other galaxies where multiple clusters were
observed, there is no observed spatial variation in outflow speed at
this resolution.  Although global winds, generally associated with
nuclear starbursts, have received the most attention to date, our work
draws attention to the need for more measurements of localized
outflows in less extreme galaxies (Schwartz 2005, PhD thesis). Hence,
the STIS slit may not have covered the star clusters with outflows in
some galaxies. Considering also that dwarf galaxy outflows are
generally slower than our detection threshold of 1-$\sigma$, the
fraction of local star-forming galaxies with cool outflows is likely
higher than our statistics (eight of 17) suggest.  

Studying galactic outflows in the ultraviolet facilitates direct 
comparison to the winds detected in high-redshift galaxies. The
continuum slope of our composite spectrum for the local star-forming 
galaxies is indistinguishable from that of an analogous rest-frame UV 
spectrum for Lyman Break Galaxies (Shapley et al. 2003). 
The constituents of the latter sample are intrinsically brighter,
but the two samples evidently have similarly young ages and average
reddening.  The low-ionization interstellar lines in these two composite 
spectra are remarkably similar with blueshifted absorption at about 
-150~\kms, which may reflect the flattening of outflow speeds seen as
a function of SFR (Martin 2005).
%and both show high-ionization interstellar \ion{Si}{4} absorption.  
The lines for the local composite spectra are broader,
reflecting 
%the inclusion of sytems with lower outflow speeds as well
a difference in spectral resolutions, as well as the exclusion of the
diffuse/field clusters in the local composite spectrum.

The case for slower outflows in dwarfs rests on a very limited number
of measurements.  The STIS G140L spectra are not well-suited
to kinematic measurements of dwarf starbursts.  We are interested in
the upper envelope of the 
velocity distribution over a broad range in star formation rates or,
better yet, rotation velocities.  Hence, the low velocities in
NGC~4214 and NGC~4670 have little impact, as do the large
uncertainties for NGC~5102.  The main results are the revised outflow
speed in He~2-10 ($-170~\pm~8$~\kms) and the new measurement for
Mrk~33 ($-200~\pm~88$~\kms). These galaxies help fill in the gap
between dwarf starburst and LIRG luminosities and are found to be
consistent with the velocity limits described by Martin (2005). The
large outflow speeds in Mrk~209 and Mrk~36 are of special interest
since these galaxies are not very luminous. Better measurements of
dynamical mass and star formation rate are needed for these galaxies. 

%__________________________________________________________________________

\acknowledgments

We thank Dr. Alice Shapley for a stimulating discussion about this
work. Financial support was provided by the David and Lucille Packard
Foundation and the Alfred P. Sloan Foundation. This research has made
use of the NASA/IPAC Extragalactic Database (NED), which is operated
by the Jet Propulsion Laboratory, California Institute of Technology,
under contract with the National Aeronautics and Space
Administration. This research has made use of the NASA Astrophysics
Data System abstract service.    

%__________________________________________________________________________

%--------------

%%% \bibitem[()]{} 

\clearpage
%--------------
%\documentstyle[preprint]{aastex}  
%\begin{document}

%\begin{deluxetable*}{lccccccccccc} 
\begin{deluxetable}{lcccccccccccc} 
\rotate
\tabletypesize{\scriptsize}
\tablecolumns{13} 
\tablewidth{0pt}
\tablecaption{{\sc The Sample of Galaxies}\label{proptab}}
\tablehead{ 
\colhead{Galaxy} & \colhead{Galaxy} 
                 & \colhead{d}   
                 & \colhead{$v_{rot}$\tablenotemark{a}}
                 & \colhead{log O/H } & \colhead{$i$} 
                 & \colhead{$M_B$} 
                 & \colhead{Age} 
                 & \colhead{$\beta$}  & \colhead{M(\ion{H}{1})} & \colhead{log} & \colhead{log}
                 & \colhead{References}\\
\colhead{}       & \colhead{Type} & \colhead{(Mpc)} 
                 & \colhead{(km s$^{-1}$)} & \colhead{+12} 
                 & \colhead{(deg)} & \colhead{} 
                 & \colhead{(Myr)} & \colhead{} & \colhead{($10^7$M$_\odot$)}
                 & \colhead{L$_{H\alpha}$} & \colhead{L$_{UV}$} & \colhead{}\\
\colhead{(1)}    & \colhead{(2)} & \colhead{(3)} 
                 & \colhead{(4)} & \colhead{(5)} 
                 & \colhead{(6)} & \colhead{(7)} 
                 & \colhead{(8)} & \colhead{(9)} 
                 & \colhead{(10)} & \colhead{(11)} 
                 & \colhead{(12)} & \colhead{(13)}
}
\startdata
He 2-10     & BCD         & 9    & 119$\pm$10 & 8.9 & ... & -17.32 &  5$\pm$1    & -1.01 & 30.5 & 40.28 & 42.74 & 1, 2, --, 1, 3          \\%& 42.74 
Mkn 33      & BCD         & 19.5 & 111$\pm$5  & 8.4 & ... & -18.31 &  5$\pm$1    & -1.84 &  62  & 41.24 & 42.61 & 4, 5, --, 6, 7          \\%& 42.61 
Mkn 36      & BCD         & 6.9  &  43$\pm$5  & 7.8 & ... & -14.02 &  $<$1       & -1.83 &   2  & 39.85 & 41.00 & 4, 8, --, 4, 7          \\%& 41.00 
Mkn 209     & BCD         & 4.9  &  47$\pm$21 & 7.8 &  30 & -13.30 &  $<$1       & -2.45 &  4.1 & 39.90 & 40.83 & 9, 8, 10, 9, 7          \\%& 40.83 
NGC 1741    & int. pair   & 51   &  ...       & 8.2 & ... & -21.15 &  5$\pm$1    & -1.35 & 1380 & 41.77 & 42.92 & --, 11, --, 13, 12      \\%& 42.92 
NGC 3125    & BCD         & 18.3 &  ...       & 8.3 & ... & -17.81 &  5$\pm$1    &  0.62 &  ... & 40.61 & 42.02 & --, 14, --, --, 7       \\%& 42.02 
NGC 3310-A\tablenotemark{b}
            & SABbc pec   & 18   & 311$\pm$10 & 8.2 &  52 & -20.13 &  5$\pm$1    & -0.98 &  588 & 41.60 & 42.73 & 14, 15, 13, 14, 16      \\%& 42.73 
NGC 3310-BC\tablenotemark{b}
            & SABbc pec   & 18   & 311$\pm$10 & 8.2 &  52 & -20.13 &  5$\pm$1    & -0.98 &  588 & 41.60 & 42.73 & 14, 15, 13, 14, 16      \\%& 42.73 
NGC 4214    & IAB(s)m     & 3.6  &  70$\pm$10 & 8.2 & ... & -17.64 &  4$\pm$1    & -1.20 &  110 & $\geq$40.40 & 41.23 & 17, 1, --, 18, 5 \\%& 41.23 
NGC 4449    & IBm         & 3.6  &  ...       & 7.8 &  51 & -17.79 &  5$\pm$1    & -2.74 &  135 & 40.65 & 41.19 & 1, 17, 17, 18, 16       \\%& 41.19 
NGC 4670    & BCD         & 16   & 110$\pm$21 & 8.2 &  28 & -17.93 &  7$\pm$1    & -1.76 &  110 & 40.76 & 42.31 & 19, 13, 20, 20, 21      \\%& 42.31 
NGC 4861    & BCD/Impec   & 10.7 &  54$\pm$3  & 7.9 &  82 & -17.25 &    ...      &  ...  &  114 & 41.09 & 42.20 & 5, 1, 6, 6, 6          \\%& 42.20 
NGC 5102    & SA0 pec     & 3.1  &  95$\pm$12 & 9.0 &  70 & -17.11 &  55$_{-37}$ &  1.50 &   34 & $>$37.93 & 40.72 & 22, 23, 22, 22, 24    \\%& 40.72 
NGC 5996    & SBc         & 47   & 142$\pm$10 & 8.9 & ... & -20.76 &  4$\pm$1    & -1.29 &  138 & 41.16 & 42.73 & 25, 13, --, 26, 27      \\%& 42.73 
NGC 6764    & SB(s)bc/Sy2 & 32   & 187$\pm$23 & 8.7 &  44 & -19.97 &  3$\pm$1    &  0.70 &  380 &  ...  & 42.03 & 28, 25, 28, 25, --      \\%& 42.03 
NGC 7552    & SB(s)ab     & 21   & 230$\pm$30 & 9.2 &  31 & -20.36 &  5$\pm$1    & -0.71 &  870 & 41.05 & 42.29 & 29, 13, 30, 31, 27      \\%& 42.29 
Tol1924-416 & BCD         & 37   &  45$\pm$10 & 8.1 & ... & -19.54 &  $<$1       & -2.67 &  300 & 41.84 & 42.95 & 32, 13, --, 32, 6       \\%& 42.95 
VII Zw 403  & BCD         & 4.5  &  20$\pm$3  & 7.7 & ... & -13.77 &    ...      &  ...  &  6.9 & 39.27 & 40.46 & 5, 8, --, 7, 7         \\%& 40.46 
\enddata
\tablerefs{(1) Sauvage et al. 1997, (2) Kobulnicky et al. 1999, (3) Johnson et al. 2000,
  (4) Bravo-Alfaro et al. 2004, (5) Davidge 1989, (6) Thuan et al. 2004, (7) Gil de Paz et al. 2003, 
  (8) P\'erez-Montero \& D\'iaz 2003, (9) Viallefond \& Thuan 1983, (10) Stil \& Israel 2002, 
  (11) Guseva et al. 2000, (12) Dopita et al. 2002, 
  (13) Heckman et al. 1998, (14) Mulder et al. 1995, (15) Pastoriza et al. 1993, 
  (16) Kennicutt et al. 1989, (17) Martin 1998, (18) Karachentsev et al. 1999, 
  (19) Hunter et al. 1996, (20) Hunter et al. 1994, (21) Marlowe et al. 1995,
  (22) van Woerden et al. 1993, (23) Chandar et al. 2004, 
  (24) McMillan et al. 1994, (25) Contini et al. 1997, (26) Kandalyan 2003, 
  (27) Buat et al. 2002, (28) Wilcots et al. 2001, (29) Lehnert \& Heckman 1996, 
  (30) Beck et al. 2002, (31) Claussen \& Sahai 1992, (32) Ostlin et al. 1998 }
\tablecomments{Col. (2): Galaxy type is from RC3.  Col. (3): Distance
  given assuming H$_o$ = 75~\kms~Mpc$^{-1}$.  Col. (4): Rotation 
  speed is from the first reference in the last column.  All rotation
  speeds are from \ion{H}{1} measurements, and are corrected for
  inclination when available.  Col. (5): 12 + log(O/H) is from the
  second reference in the last column.  Col. (6):  Inclination is from
  third reference in the last column.  Col. (7): Absolute blue
  magnitude $M_B$ is calculated from the RC3 magnitude $B_T^0$.
  Col. (8): \ion{H}{1} mass is from fourth reference in the final column.
  Col. (9): Age is from Chandar et al. (2004). Age is based on
  the comparison of STIS FUV spectra with single burst STARBURST99
  population synthesis models (Leitherer 1999).  Col. (10): Power-law index of the UV
  continuum (F~$\propto\lambda^\beta$) in the STIS G140L spectra over
  the wavelength range 1240 - 1600~\AA~after correcting for foreground
  Milky Way reddening. $\beta$ is measured to an accuracy of $\pm
  0.01$. (From Chandar et al. 2004.) Col. (11): H$\alpha$ luminosity in ergs s$^{-1}$
  is from the fifth/last reference in the last column; these values
  are corrected for Galactic extinction and [NII]
  ($\lambda\lambda$6548, 6583) contamination, but not for internal 
  reddening.  Col. (12): UV luminosity in ergs s$^{-1}$ is from {\em IUE} flux at 1900~\AA, 
  except He~2-10 and NGC 1741, where L$_{UV}$ is $\lambda$L$_\lambda$ at 1500~\AA,
  from Johnson et al. (2000) and Conti et al. (1996), respectively. To account for the UV spectral slope, $\beta$
  (Col. 9), we followed Bell \& Kennicutt (2001).}
\tablenotetext{a}{Rotation speeds are from \ion{H}{1} rotation curves for
  NGC~3310, NGC~4214, NGC~4670, NGC~5102, and NGC~7552.  All other
  rotation speeds are from $\Delta$v$_{50}$, the \ion{H}{1} velocity
  width at the 50\% level.}
\tablenotetext{b}{NGC 3310-ABC are \ion{H}{2} regions within NGC
  3310.  NGC 3310-A (see Figure \ref{3310map}) is
  equivalent to NGC 3310-1 in Chandar et al. (2004), while NGC 3310-B
  and -C are part of NGC 3310-all in that paper. }
%\footnote{
%  The {\em IUE} UV spectra from Kinney et al. (1993) were used to measure
%  the  UV flux for most of the galaxies. The data for 
%  He~2-10 and NGC~1741 were taken from Johnson et al. (2000) and Conti et al. 
%  (1996), respectively.  To account for the UV spectral slope, $\beta$
%  (column 9 of Table~\ref{proptab}), we followed Bell \& Kennicutt
%  (2001).} 

\end{deluxetable}

%\end{deluxetable*}

%\end{document}

\clearpage
%--------------
%\documentstyle[preprint]{aastex}  
%\begin{document}

%\begin{deluxetable*}{lcccccc} 
\begin{deluxetable}{lclccccccc} 
\tabletypesize{\scriptsize}
%\rotate
\tablecolumns{10} 
\tablewidth{0pt}
\tablecaption{{\sc Absorption Line Properties}\label{kintab}}
\tablehead{ 
\colhead{Galaxy} & \colhead{No. of} & \colhead{Spatial}
                 & \colhead{$v_{sys}$}
                 & \colhead{$v_{sys}$}
                 & \colhead{$\Delta v_{outflow}$}
                 & \colhead{FWHM$_{inst}$}
                 & \colhead{FWHM(\ion{C}{2})} & \colhead{EW(\ion{C}{2})} & \colhead{} \\
\colhead{}       & \colhead{Clusters} & \colhead{Extent}
                 & \colhead{(km s$^{-1}$)}
                 & \colhead{source\tablenotemark{a}}
                 & \colhead{(km s$^{-1}$)}
                 & \colhead{(\AA)\tablenotemark{b}}
                 & \colhead{(km s$^{-1}$)\tablenotemark{c}} & \colhead{(\AA)}
                 & \colhead{Reference}\\
\colhead{(1)}    & \colhead{(2)} & \colhead{(3)} 
                 & \colhead{(4)} & \colhead{(5)} 
                 & \colhead{(6)} & \colhead{(7)} 
                 & \colhead{(8)} & \colhead{(9)} 
                 & \colhead{(10)} 
}
\startdata
He 2-10      & 5 & +4\farcs5, -1\farcs2 &  869$\pm$3   & \ion{Mg}{1} & -170$\pm$8   & 2.5 & 1136 & 5.0 & 1  \\%& 1272/572 \\ OUTFLOW
Mkn 33       & 2 & -0\farcs9, -4\farcs1 & 1465$\pm$10  &  opt. lines & -200$\pm$88  & 3.8 &  555 & 2.4 & 2  \\%& 1015/850 \\ OUTFLOW
Mkn 36       & 6 & -0\farcs6, -1\farcs3 &  646$\pm$5   &  \ion{H}{1} & -213$\pm$68  & 2.6 & 1203 & 3.4 & 3  \\%& 1338/586 \\ OUTFLOW
Mkn 209      & 3 & +2\farcs0, -2\farcs2 &  288$\pm$15  &          CO & -191$\pm$92  & 1.6 &  928 & 1.7 & 4  \\%& 1397/359 \\ OUTFLOW
NGC 1741     & 4 & +2\farcs7, -3\farcs1 & 3937$\pm$45  &  opt. lines &   16$\pm$76  & 3.2 & 345  & 2.1 & 5  \\%&  795/716 \\ NO OUTFLOW
NGC 3125     & 1 & -0\farcs9, -1\farcs2 &  964$\pm$73  &    UV lines &    8$\pm$73  & 2.5 &  445 & 1.1 & 6  \\%&  717/562 \\ NO OUTFLOW
NGC 3310-A\tablenotemark{d}
             & 1 & -4\farcs3, -4\farcs6 & 1040$\pm$12  &  \ion{H}{1} & -339$\pm$72  & 2.9 & 1088 & 4.5 & 7  \\%& 1268/652 \\ OUTFLOW
NGC 3310-B\tablenotemark{d}
             & 1 & +8\farcs5, +8\farcs1 & 1010$\pm$12  &  \ion{H}{1} & -414$\pm$82  & 2.9 & 1278 & 5.1 & 7  \\%& 1435/652 \\ OUTFLOW
NGC 3310-C\tablenotemark{d}
             & 1 & +7\farcs2, +7\farcs0 & 1010$\pm$12  &  \ion{H}{1} & -561$\pm$83  & 1.9 & 1382 & 4.7 & 7  \\%& 1446/427 \\ OUTFLOW
NGC 4214     & 1 & -0\farcs8, -1\farcs3 &  304$\pm$1   & \ion{Mg}{1} &  -32$\pm$49  & 3.3 &  549 & 2.1 & 8  \\%&  923/742 \\ NO OUTFLOW
NGC 4670     & 2 & +2\farcs5, -1\farcs6 & 1070$\pm$21  &  \ion{H}{1} &  -65$\pm$73  & 3.3 &  527 & 1.6 & 9  \\%&  910/742 \\ OUTFLOW
NGC 5102     & 2 & -0\farcs9, -1\farcs2 &  470$\pm$12  &  \ion{H}{1} &  -98$\pm$78  & 2.3 &  859 & 3.3 & 10 \\%& 1110/517 \\ OUTFLOW
NGC 5996     & 2 & -0\farcs8, -2\farcs9 & 3297$\pm$32  &  \ion{H}{1} & -289$\pm$59  & 2.7 &  735 & 2.7 & 3  \\%&  957/613 \\ OUTFLOW
NGC 6764     & 1 & -0\farcs8, -1\farcs2 & 2420$\pm$20  &          CO &   -9$\pm$50  & 2.6 &  653 & 3.1 & 11 \\%&  876/584 \\ NO OUTFLOW
NGC 7552     & 2 & +1\farcs7, +0\farcs9 & 1457$\pm$108 &    UV lines & -316$\pm$108 & 3.5 &  856 & 5.6 & 5  \\%& 1163/787 \\ OUTFLOW
Tol 1924-416 & 2 & +4\farcs6, -1\farcs6 & 2935$\pm$70  &    UV lines &  -47$\pm$70  & 2.3 &  391 & 1.1 & 10 \\%&  640/517 \\ OUTFLOW 
%
% not-de-convolved line widths are commented out.  line widths are
% deconvolved assuming a filled slit, i.e. 4.67 Anstroms.
%
\enddata
\tablerefs{(1) Schwartz \& Martin 2005 (in prep), (2) Lequeux et
  al. 1995, (3) RC3, (4) Young et al. 1995, (5) Vacca \& Conti
  1992, (6) this paper, (7) Mulder et al. 1995, 
  (8) Becker et al. 1995, (9) Hunter \& Hoffman 1999, (10) van
  Woerden et al. 1993, (11) Eckart et al. 1996}  
\tablecomments{The number of clusters given in Col. (2) is the number 
  of individual cluster spectra extracted and averaged to produce the 
  final spectrum of the region.  The spatial extent given in Col. (3) 
  is the maximum distance from the center of the slit to the clusters 
  both above and below the center of the slit, in arcseconds.   
  $\Delta v_{outflow}$ = $v-v_{sys}$ given in Col. (6)  is an average
  of all detectable ($\ge 3\sigma$) interstellar absorption lines.  
  Since \ion{C}{2} is the only unblended line detectable in every spectrum, 
  its Gaussian FWHM is given in Col. (8), as it is believed to be 
  representative of the typical absorption line.  NGC 4449, NGC 4861, and 
  VII Zw 403 were left out of this table due to very low ($\lesssim$1) S/N.}
\tablenotetext{a}{Source of $v_{sys}$ is given as one of the
  following: \ion{Mg}{1} b-band
  (stellar) absorption lines, optical emission lines,
  \ion{H}{1} 21-cm, CO lines, or UV absorption lines.}
\tablenotetext{b}{The instrumental line width is measured from the
  two-dimensional spectra, as described in \S\ref{resn}.}
\tablenotetext{c}{The ``actual'' FWHM of \ion{C}{2} is measured from the
  spectra and deconvolved in quadrature from the instrumental width,
  as described in \S\ref{resn}.}   
\tablenotetext{d}{Systemic velocities for NGC 3310 are from \ion{H}{1}
  velocity field maps from Mulder et al. (1995) to account for
  rotation, and show that the outflow velocities in different
  star-forming regions of NGC 3310 are certainly not identical.}
%\tablenotetext{e}{For these four galaxies, the instrumental line width
%  measured is consistent with the width of foreground Galactic/Milky
%  Way Halo lines measured.  No other galaxies showed foreground
%  absorption lines in their spectra.}\tablenotemark{e}

\end{deluxetable}

%\end{deluxetable*}

%\end{document}

\clearpage
%--------------
%\documentstyle[preprint]{aastex}  
%\begin{document}

%\begin{deluxetable*}{lcccccl} 
\begin{deluxetable}{lcccl} 
\tabletypesize{\scriptsize}
\tablecolumns{6} 
\tablewidth{0pt}
\tablecaption{{\sc Absorption Lines of Interest}\label{linetable}}
\tablehead{ 
\colhead{Line} & \colhead{$\lambda$ (lab)\tablenotemark{a}}   
                 & \colhead{I.E.\tablenotemark{b}}
            %     & \colhead{I.E.(2)\tablenotemark{b}}
                 & \colhead{Astrophysical} & \colhead{Notes}\\%
                 & \colhead{(\AA)} 
                 & \colhead{(eV)} %& \colhead{(eV)}
                 & \colhead{Source\tablenotemark{c}} & \colhead{}
}
\startdata
\ion{C}{3}  & 1175.65 & 24.38  & Stellar &   \\%& 47.89
\ion{N}{5}  & 1238.80 & 77.47  & Blend\tablenotemark{d} & Blends with
Ly~$\alpha$ absorption\\%& 97.89
            & 1242.78 & 77.47  & Blend\tablenotemark{d} & Blends with
Ly~$\alpha$ absorption\\%& 97.89
\ion{C}{3}  & 1247.38 & 24.38  & Stellar & Blends with \ion{Si}{2} $\lambda$1250 \\%& 47.89
\ion{Si}{2} & 1260.42 &  8.15  &     ISM & Blends with \ion{C}{3} $\lambda$1247 \\%& 16.35 
\ion{O}{1}\tablenotemark{e}   
            & 1302.17 &  0.00  & ISM & Blends with \ion{Si}{2} $\lambda$1304\\%& 13.62
\ion{Si}{2}\tablenotemark{e}  
            & 1304.37 &  8.15  & ISM & Blends with \ion{O}{1} $\lambda$1302\\%& 16.35
%\ion{C}{2} & 1323.95 & 11.26  & Stellar &   \\%& 24.38 
\ion{C}{2}  & 1334.53 & 11.26  &     ISM &   \\%& 24.38
\ion{Si}{3} & 1343.41 & 16.35  & Stellar &   \\%& 33.49
\ion{Si}{4} & 1393.75 & 33.49  &   Blend\tablenotemark{f} & Often shows PCyg structure \\%& 45.14 
            & 1402.77 & 33.49  &   Blend\tablenotemark{f} & Often shows PCyg structure \\%& 45.14
\ion{Si}{3} & 1417.24 & 16.35  & Stellar & Blends with \ion{Si}{4} in PCyg profiles \\%& 33.49
\ion{C}{3}  & 1426.80 & 24.38  & Stellar &   \\%& 47.89
\ion{Fe}{5} & 1430.57 & 54.80  & Stellar &   \\%& 75.0 
\ion{S}{5}  & 1501.76 & 47.22  & Stellar &   \\%& 72.59
\ion{Si}{2} & 1526.71 &  8.15  &     ISM & Blended with \ion{C}{4} \\%& 16.35 
\ion{C}{4}  & 1548.19 & 47.98  &   Blend\tablenotemark{f} & Often shows PCyg structure \\%& 64.49
            & 1550.77 & 47.98  &   Blend\tablenotemark{f} & Often shows PCyg structure \\%& 64.49
\ion{Fe}{2} & 1608.45 &  7.87  &     ISM &   \\%& 16.19
\ion{He}{2} & 1640.34 & 24.59  & Stellar &   \\%& 54.42
\ion{Al}{2} & 1670.78 &  5.99  &     ISM &   \\%& 18.83
\ion{Fe}{2} & 2344.21 &  7.87  &     ISM &   \\%& 16.19
\ion{Fe}{2} & 2382.77 &  7.87  &     ISM &   \\%& 16.19
\ion{Fe}{2} & 2586.65 &  7.87  &     ISM &   \\%& 16.19
\ion{Fe}{2} & 2600.17 &  7.87  &     ISM &   \\%& 16.19
\ion{Mg}{2} & 2796.35 &  7.65  &     ISM\tablenotemark{g} &   \\%& 15.04
            & 2803.53 &  7.65  &     ISM\tablenotemark{g} &   \\%& 15.04
\ion{Mg}{1} & 2852.96 &  0.00  &   Blend\tablenotemark{g} &   \\%&  7.65
\\%
\ion{Na}{1} & 5889.95 &  0.00  &     ISM\tablenotemark{h} &   \\%&  5.14
            & 5895.92 &  0.00  &     ISM\tablenotemark{h} &   \\% &  5.14
\enddata
\tablenotetext{a}{All lab (air) wavelengths are from V\'azquez et al. 2004
  except: \ion{C}{3} $\lambda$1427, \ion{Fe}{5} $\lambda$1431, and
  \ion{He}{2} $\lambda$1640 (from Morton 1991); \ion{Si}{2}
  $\lambda$1260 and \ion{Fe}{2} $\lambda$1608 (from Shapley et al. 2003).}
\tablenotetext{b}{The energy given as ``I.E.'' is the energy required
  to remove an electron from the parent ion to create the ion listed.
  For example, I.E.(\ion{Al}{2}) is the energy needed to remove an
  electron from neutral Al and create the \ion{Al}{2} ion. Values are
  from Cox 2000.} 
\tablenotetext{c}{Possible sources are predominantly stellar
  (``Stellar''), predominantly interstellar (``ISM''), or a blend
  (``Blend'').} 
\tablenotetext{d}{\ion{N}{5} $\lambda\lambda$~1238, 1242 is severely
  blended with Ly~$\alpha$ absorption in nearly all galaxies, as is
  also the case with the Lyman Break Galaxies (Shapley et al. 2003).}
\tablenotetext{e}{\ion{O}{1} and \ion{Si}{2} are blended together at
  our spectral resolution.  We use the same average wavelength as
  Shapley et al. 2003: 1303.27 \AA.}
\tablenotetext{f}{\ion{Si}{4} and \ion{C}{4} are assumed to be blends
  of a P Cygni emission/absorption profile, absorption in stellar
  photospheres, and possibly some interstellar absorption 
  (e.g., Heckman et al. 1998; Shapley et al. 2003; Chandar et al. 2004).}  
\tablenotetext{g}{\ion{Mg}{2} is assumed to be predominantly interstellar 
  based on the findings of V\'azquez et al. 2004 wherein no stellar components 
  were observed for these ions in a high-resolution spectrum of a nearby 
  dwarf starburst. \ion{Mg}{1} is present as an interstellar line in NGC~1705 
  (V\'azquez et al. 2004), but is also commonly seen in stellar UV spectra 
  (Fanelli et al. 1992).  Since the optical \ion{Mg}{1} triplet (5172~\AA) 
  is stellar as well, we assume this line as a blend.}
\tablenotetext{h}{Data for the optical resonance doublet, \ion{Na}{1},
  also called the Na D doublet, is provided for comparison.  Schwartz
  \& Martin (2004; 2005, in prep) analyze the Na D absorption in NGC
  4214, NGC 4449 and He2-10, as discussed in the text.}
\end{deluxetable}
%\end{deluxetable*}

%\ion{}{}  &  &  &      &   \\%

%\end{document}

%--------------
%\input{f1.tex}
\clearpage
\begin{figure}
\epsscale{0.9}
%\plotone{f1.eps}
\caption{Uncertainty ($\delta v$) in velocity (km s$^{-1}$) as a function of
  median signal-to-noise for the spectra fit with
  SPECFIT: NGC 3310-ABC, NGC 4214, NGC 7552, and He 2-10.  (Fitting a
  spectrum with SPLOT does not give a useful measurement of error.)
  The data are fit (dashed line) with a least-squares fit, giving a
  slope -1.8~\kms~$\pm$~0.2~\kms per unit signal-to-noise. \label{snrerror}}  
\end{figure}
\clearpage
%--------------
%\input{f2.tex}
\begin{figure}
\epsscale{0.5}
%\plotone{f2a.eps}
%\plotone{f2b.eps}
\caption{(a) Normalized, rest-frame STIS G140L spectra of the \ion{C}{2} 
  absorption line. The rest wavelength of \ion{C}{2} (1334.53~\AA)
  is marked by the dotted line.  This is one of the only unblended
  interstellar lines which appears in all spectra. The top three rows
  shows \ion{C}{2} in all the galaxies with outflows; the bottom row
  shows \ion{C}{2} in the four galaxies where little or no outflow is
  detected. 
  (b) The same plot, but for the \ion{Si}{2} absorption line at
  1260.42~\AA.  The two lines plotted illustrate the variation in
  velocity between lines.  For example, some galaxies (e.g., He~2-10)
  exhibit obviously blueshifted lines with similar profiles in both
  \ion{C}{2} and \ion{Si}{2}, whereas other galaxies (e.g., NGC~4670)
  show differences between the two lines. \label{ciifig}} 
\end{figure}
\clearpage
%--------------
%\input{f3.tex}
\begin{figure}
\epsscale{0.3}
%\epsscale{1.0}
%\plotone{f3a_color.eps}
%\plotone{f3b_color.eps}
\caption{(a) Normalized STIS G140L spectra of the low ionization
  absorption line profiles used to measure the warm/cold gas outflow
  velocity in He 2-10.  Top: \ion{O}{1}/\ion{Si}{2} $\lambda$1303 and
  \ion{C}{2} $\lambda$1335. Bottom: \ion{Si}{2} $\lambda$1263 and
  \ion{Al}{2} $\lambda$1671.  The solid line plotted over the data is the
  fit from the SPECFIT program.  The dotted lines represent the
  wavelengths corresponding to the systemic velocity for each of the
  absorption lines, which are labelled with the species and ionization
  state. The systemic velocity used is 869 \kms, as discussed in
  the text; this is consistent with the stellar lines seen in this
  spectrum. The interstellar absorption is measured to have $\Delta
  v_{outflow}$ = -170~\kms.  
  (b) Top: Same as (a), for G230L grating lines \ion{Fe}{2}
  $\lambda\lambda$2344, 2374, 2383, \ion{Mg}{2}
  $\lambda\lambda$2796, 2804, and \ion{Mg}{1} $\lambda$2853.
  Bottom: Keck/HIRES spectra ($\sim$7~\kms~resolution) of Na D
  $\lambda\lambda$5890, 5896. The interstellar \ion{Mg}{1} and \ion{Mg}{2} 
  lines are fit to have the same blueshift of -170~\kms~as the optical
  absorption lines (Schwartz \& Martin 2006, in prep.); the optical
  lines are statistically well-fit ($\chi^2$ = 0.75).  The \ion{Fe}{2}
  lines show smaller blueshifts, if any at all; these are blends of
  multiplet lines and nearby absorption in the Galactic halo and
  clouds.\label{he210}} 
\end{figure}
\clearpage

%--------------
%\input{f4.tex}
\begin{figure}
\epsscale{0.9}
%\plotone{f4.jpg}
\caption{Archival {\em HST}/WFPC2 F606W (MAST dataset \#U2E67Q01T)
  image of the central region of NGC 3310, with position of the STIS
  slit overlaid.  The star-forming regions A, B, and C are marked.
  Regions A and C are $\sim$850 pc apart, and regions B and C are
  separated by roughly 100 pc. \label{3310map}} 
\end{figure}
\clearpage
%--------------
%\input{f5.tex}
\begin{figure}
\epsscale{0.9}
%\plotone{f5_color.eps}
\caption{STIS G140L spectra of the three bright star clusters (A, B,
  and C) observed in NGC~3310.  The spectra have been
  de-redshifted. The solid lines (colored blue in the electronic
  edition) represent the wavelengths of stellar lines, and they are
  \ion{Si}{3} $\lambda$1417, \ion{C}{3} $\lambda$1427, and \ion{S}{5}
  $\lambda$1502.  These are fit to check the value of the systemic
  velocity used.  The dashed (red) lines are predominantly
  interstellar in origin; they are \ion{Si}{2} $\lambda$1260,
  \ion{O}{1}/\ion{Si}{2} $\lambda$1303, \ion{C}{2} $\lambda$1335,
  \ion{Si}{2} $\lambda$1527, and \ion{Al}{2} $\lambda$1671. These are
  used to fit the velocity of the gas in the interstellar medium.  The
  dotted (green) lines are the \ion{Si}{4} and \ion{C}{4} doublets,
  which are blends of the gas in stellar photospheres, winds, and
  possibly some interstellar gas. \label{diskspec} } 
\end{figure}
\clearpage

%--------------

%\input{f6.tex}
\begin{figure}
\epsscale{0.9}
%\plotone{f6_color.eps}
\caption{Normalized STIS spectra of the low ionization absorption line
  profiles used to measure the interstellar gas outflow velocity in NGC
  7552.  The solid line plotted over the data is the fit from the 
  SPECFIT program.  The dotted lines represent the wavelengths
  corresponding to the systemic velocity for each of the absorption
  lines, which are labelled with the species and ionization state.
  The systemic velocity is 1457~\kms~(from stellar absorption lines in
  the G140L spectrum); this is consistent with the stellar lines seen
  in this spectrum.  The average outflow velocity of cold/warm gas as
  measured by low ionization lines such as \ion{C}{2}, \ion{Si}{2},
  and \ion{Al}{2} is 316~\kms.\label{7552fig}} 
\end{figure}
\clearpage
%--------------
%\input{f7.tex}
\begin{figure}
\epsscale{0.9}
%\plotone{f7_color.eps}
\caption{Average outflow velocity ($\Delta v$) for all interstellar
  absorption lines as a function of the FWHM of the \ion{C}{2} line,
  which is often the only detectable, unblended, purely interstellar
  absorption line. The error bars are estimated from the relation
  between velocity fitting errors and signal-to-noise as discussed in
  \ref{dopps}.  The solid line (black in the electronic edition) is a
  least-squares fit of the data which shows that FWHM $\sim 2\Delta
  v_{outflow}$ (actual slope of fit = $1.6\pm 0.50$), as discussed in
  the text.  
%The dashed line (blue) represents FWHM = $2\Delta v_{outflow}$
\label{fwhmfig}}     
\end{figure}
\clearpage
%--------------
%\input{f8.tex}
\begin{figure}
\epsscale{0.9}
%\plotone{f8_color.eps}
\caption{ Dependence of starburst region age (top), UV continuum slope
  $\beta_i$ (middle), and oxygen abundance log(O/H) + 12 (bottom) on outflow velocity.
  Age and $\beta_i$ are taken from models calculated in Chandar et
  al. (2004); NGC 5102 has an age of 55$_{-37}$, and is not included
  in this graph for the sake of clarity.  Oxygen abundances given as
  log(O/H) + 12 are from references given in Table \ref{proptab};
  abundance errors are not specified.  The given outflow velocities
  are averages of all available, unblended interstellar absorption
  lines (as discussed in the text). Velocity uncertainties are
  estimated using the signal-to-noise of the individual spectra as
  discussed in Section \ref{dopps}, and are given in Table
  \ref{kintab}. The filled triangles (colored red in the electronic
  edition) represent dwarf galaxies; the (green) filled squares
  represent Mkn~33, NGC~5102, and Tol~1924-416, which aren't
  classified as dwarfs or disks in this paper; (blue) open stars
  represent disk galaxies. \label{params}}   
\end{figure}
\clearpage
%--------------
%\input{f9.tex}
\begin{figure}
\epsscale{0.9}
%\plotone{f9_color.eps}
\caption{Outflow velocity ($\Delta v_{outflow}$ = $v_{line} -
  v_{sys}$; from an average of all available interstellar absorption
  lines) vs. galaxy rotation speed for the sources showing outflows.
  Disk galaxies (NGC~3310-ABC, NGC~5996, NGC~7552) are represented by
  stars (blue in the electronic edition); galaxies which are dwarfs by
  morphology but not by M$_B$ (Mkn~33, NGC~5102, Tol~1924-416) are
  represented by a filled square (green). Dwarf galaxies are marked by
  filled triangles (red).  NGC~1705 is marked by a filled circle
  (cyan; V\'azquez et al. 2004).  The error bars from the STIS data
  represent the velocity fitting error as discussed in \S \ref{dopps}.
  The dotted (red) diagonal lines show the relations $v_{term} =
  \sqrt{2} v_{rot}$ (minimum escape velocity) and $3 v_{rot}$
  (escape velocity at 3~kpc in an isothermal halo extending to
  100 kpc).  The solid (black) line shows a linear least squares fit
  to the data; its slope is 1.2 $\pm$ 0.2. \label{vtvr}}   
\end{figure}
\clearpage
%--------------
%\input{f10.tex}
\begin{figure}
\epsscale{0.9}
%\plotone{f10_color.eps}
\caption{\ion{C}{2} outflow velocity versus \ion{Al}{2} (top) and
  \ion{Si}{2} (bottom) outflow velocities.  The ionization energy of
  carbon is 11.26 eV, whereas the ionization energy of aluminum
  (silicon) is 5.99 eV (8.15 eV).  The diagonal (dotted) lines
  represent $v_{CII} = v_{AlII}$ and $v_{CII} = v_{SiII}$.  The
  least-squares fits of the lines give slopes of $0.77\pm 0.15$
  ($0.61\pm 0.13$) for \ion{Al}{2} (\ion{Si}{2}).
  \label{cvsal}}  
\end{figure}
\clearpage
%--------------
%\input{f11.tex}
\begin{figure}
\epsscale{0.9}
%\plotone{f11_color.eps}
\caption{(a) Normalized STIS spectra of the low ionization absorption line
  profiles used to measure the warm/cold gas outflow velocity in
  NGC~4214, Region 1.  The solid (red in the electronic edition) line
  plotted over the data is the fit from the SPECFIT program.  The
  dotted lines represent the wavelengths corresponding to the systemic
  velocity for each of the absorption lines, which are labelled with
  the species and ionization state. The systemic velocity is
  304~\kms~(Becker et al. 1995), which agrees with stellar lines
  seen in optical (Schwartz \& Martin 2004) and UV (this paper).  The
  measured line velocity is $-32\pm 49$~\kms, which is consistent with
  the findings of Schwartz \& Martin 2004 that no outflow is detected
  in optical resonance absorption lines in Region 1.  (b) Normalized
  \ion{Na}{1} Keck I/HIRES optical resonance absorption spectrum for
  NGC~4214-1 (Schwartz \& Martin 2004).  The doublet lines are marked
  D1 ($\lambda$5889.95) and D2 ($\lambda$5895.92).  (c) Same as (b),
  for NGC~4214-2.  Cold, interstellar gas is seen in absorption and is
  blueshifted from the systemic velocity by 23 \kms.\label{n4214fig}}  
\end{figure}
\clearpage
%--------------
%\input{f12.tex}
\begin{figure}
\epsscale{0.9}
%\plotone{f12_color.eps}
\caption{Outflow velocity ($\Delta v_{outflow}$ = $v_{line} -
  v_{sys}$; from an average of all available interstellar absorption
  lines) vs. star formation rate as calculated from H$\alpha$
  luminosity using equation 2.  
%galaxy UV + FIR luminosity for the sources showing outflows.
  The error bars represent the velocity fitting error as discussed in
  \S \ref{dopps}.  Data point symbols and colors are the same as in
  Figure~\ref{vtvr}.  The H$\alpha$ luminosities have not been
  corrected for internal extinction, which would increase the
  luminosity (and therefore the SFR) by $\sim$50\%.  The data point with the
  lowest SFR, NGC~5102, is a lower limit on SFR because the \ha
  luminosity is that of the nucleus only. The solid line is a
  least-squares fit to the data (excepting NGC~5102); the slope is
  $0.51 \pm 0.59$. \label{sfrfig}}   
\end{figure}

%--------------
%\input{f13.tex}
\begin{figure}
\epsscale{0.9}
%\plotone{f13_color.eps}
\caption{(Top) Composite rest-frame UV spectrum of 811 Lyman Break
  Galaxies from Shapley et al. (2003); resolution $\sim$2.8~\AA.
  (Bottom) Composite rest-frame UV spectrum of the 16 star-forming
  clusters in our sample; resolution $\sim$2.9~\AA. The solid (colored
  blue in the electronic edition) lines represent the wavelengths of
  stellar lines, the dashed (red) lines are are predominantly
  interstellar in origin, and the dotted (green) lines are the
  high-ionization \ion{Si}{4} and \ion{C}{4} doublets, which are
  blends of the gas in stellar photospheres/winds, and possibly some
  interstellar gas. \label{lbgfig}}      
\end{figure}
\clearpage

%--------------
%\input{f14.tex}
\begin{figure}
\epsscale{0.9}
%\plotone{f14.jpg}
\caption{Distribution of UV luminosities (\luv; ergs s$^{-1}$
  \AA$^{-1}$) for the local sample (\luv $<10^{40}$; solid bold line
  -- red in the electronic edition) and the LBG sample (log \luv
  $>10^{40}$; Shapley et al. 2003; dashed line -- blue in the
  electronic edition).\label{luvhist}} 
\end{figure}

%__________________________________________________________________________


\begin{thebibliography}{}

\bibitem[Beck et al.(2002)]{beck02} Beck, R., et al. 2002, A\&A, 391,
  83 
\bibitem[Becker et al. 1995]{beck95} Becker, R., Henkel, C., Bomans,
  D. J., \& Wilson, T. L. 1995, A\&A, 295, 302
\bibitem[Bell \& Kennicutt(2001)]{bk01} Bell, E. F. \& Kennicutt,
  R. C. 2001, ApJ, 548, 681
\bibitem[Binney \& Tremaine(1987)]{bt} Binney, J., \& Tremaine, S. 1987,
  Galactic Dynamics (Princeton: Princeton Univ. Press)
\bibitem[Bravo-Alfaro et al.(2004)]{ba04} Bravo-Alfaro, H., Brinks,
  E., Baker, A. J., Walter, F., \& Kunth, D. 2004, AJ, 127, 264
\bibitem[Buat et al.(2002)]{buat} Buat, V., Boselli, A., Gavazzi, B., \&
  Bonfanti, C. 2002, A\&A, 383, 801
%\bibitem[Campbell et al.(1986)]{ctm86} Campbell, A., Terlevich, R., \&
%  Melnick, J. 1986, MNRAS, 223, 811
\bibitem[Chandar et al.(2003)]{rupali03} Chandar, R., Leitherer, C.,
  Tremonti, C. A., \& Calzetti, D. 2003, 586, 939
\bibitem[Chandar et al.(2004)]{rupali04} Chandar, R., Leitherer, C., \&
  Tremonti, C. A. 2004, ApJ, 604, 153
\bibitem[Conti et al. (1996)]{conti} Conti, P. S., Leitherer, C., \&
  Vacca, W. 1996, ApJ, 461, L87 
\bibitem[Contini(1996)]{cont96} Contini, T. 1996, Ph.D. Thesis, Universite
  Paul Sabatier, Toulouse, France  
%\bibitem[Contini et al.(1997)]{cont97} Contini, T., Wozniak, H.,
%  Considere, S., \& Davoust, E. 1997, A\&A, 324, 41
\bibitem[Claussen \& Sahai(1992)]{cs92} Claussen, M. J. \& Sahai,
  R. 1992, AJ, 103, 1134
\bibitem[Cox(2000)]{aaq} Cox, A. N. 2000, New York: AIP Press, Springer
\bibitem[Dahlem et al.(1998)]{dahl98} Dahlem, M., Weaver, K. A., \& Heckman,
  T. M. 1998, ApJS, 118, 401 
\bibitem[Davidge(1989)]{dav89} Davidge, T. J. 1989, PASP, 101, 494 
\bibitem[Dekel \& Silk(1986)]{dk86} Dekel, A., \& Silk, J. 1986, ApJ,
  303, 39 
\bibitem[Dopita et al.(2002)]{dopita} Dopita, M. A., Pereira, M.,
  Kewley, L. J., \& Capaccioli, M. 2002, ApJS, 143, 47
\bibitem[Eckart et al.(1996)]{eck96} Eckart, A., et al. 1996, ApJ,
  472, 588
\bibitem[Fanelli et al.(1992)]{fan} Fanelli, M. N., O'Connell, R. W.,
  Burstein, D., \& Wu, C. 1992, ApJS, 82, 197
%\bibitem[Fouque et al.(1992)]{fouque} Fouque, P. et al. 1992,
%  Catalogue of Radial Velocities
%\bibitem[Gonzalez-Delgado et al. 1998]{gd98} Gonzalez-Delgado, R.,
%  Leitherer, C., Heckman, T., Lowenthal, J., Ferguson, H., \& Robert,
%  C. 1998, ApJ, 495, 698
%\bibitem[Grevesse \& Sauval(1998)]{gs98} Grevesse, N. \& Sauval,
%  A. J. 1998, Space Sci. Rev., 85, 161
\bibitem[Gil de Paz et al.(2003)]{gdp} Gil de Paz, A., Madore, B. F.,
  \& Pevunova, O. 2003, ApJS, 147, 29
\bibitem[Guseva et al.(2000)]{gus00} Guseva, N., Izotov, Y. I., Thuan,
  T. X. 2000, ApJ, 531, 776
\bibitem[Heckman \& Leitherer(1997)]{hl97} Heckman, T. M. \&
  Leitherer, C. 1997, AJ, 114, 69
\bibitem[Heckman(1998)]{h98} Heckman, T. M. 1998, in ASP
  Conf. Ser. 148, Origins, ed. C. E. Woodward, J. M. Shull, \&
  H. A. Thronson (San Francisco: ASP), 127
\bibitem[Heckman et  al.(1998)]{heck98} Heckman, T. M., Robert, C., 
  Leitherer, C., Garnett, D. R., \& van der Rydt, R. 1998, ApJ, 503,
  646 
\bibitem[Heckman et al.(2000)]{hlsa} Heckman, T., Lehnert, M.,
  Strickland, D., \& Armus, L. 2000, ApJS, 129, 493
%\bibitem[Huchtmeier et al.(2003)]{hkk} Huchtmeier, W. K.,
%  Karachentsev, I. D., \& Karachentseva, V. E. 2003, A\&A, 401, 483
\bibitem[Hunter et al.(1994)]{hunt94} 	Hunter, D.~A., van
  Woerden, H., \& Gallagher, J.~S. 1994, ApJS, 91, 79
\bibitem[Hunter et al.(1996)]{hunt96} Hunter, D. A., van Woerden, H.,
  \& Gallagher, J. S. 1996, ApJS, 107, 739
\bibitem[Hunter \& Hoffman(1999)]{hh99} Hunter, D. A. \& Hoffman,
  L. 1999, AJ, 117, 2789
\bibitem[Johnson et al.(2000)]{kj00} Johnson, K. E., Leitherer, C.,
  Vacca, W. D., \& Conti, P. S. 2000, ApJ, 120, 1273
\bibitem[Kandalyan(2003)]{kan03} Kandalyan, R. A. 2003, A\&A, 404, 513
\bibitem[Karachentsev(1999)]{kmh99} Karachentsev, I. D., Makarov, D. I., \& Huchtmeier, W. K. 1999, A\&AS, 139, 97
\bibitem[Kennicutt(1989)]{k89.1} Kennicutt, R. C. 1989, ApJ, 344, 685
\bibitem[Kennicutt et al.(1989)]{k98.2} Kennicutt, R. C., Edgar,
  K. B., \&  Hodge, P. W. 1989, ApJ, 337, 761
\bibitem[Kinney et al.(1993)]{kinn} Kinney, A. L., Bohlin, R. C.,
  Calzetti, D., Panagia, N., \& Wyse, R. F. G. 1993, ApJS, 86, 5
\bibitem[Kobulnicky et al.(1999)]{kkp99} Kobulnicky, H. A., Kennicutt,
  R. C., \& Pizagno, J. L. 1999, ApJ, 514, 544
\bibitem[Kriss(1994)]{krisskross} Kriss, G. 1994, ASP Conf. Ser. 61,
  in Astronomical Data Analysis Software and Systems III,
  ed. D. R. Crabtree, R. J. Hanisch, \& J. Barnes (San Francisco:
  ASP), 437 
\bibitem[Kunth et al.(1998)]{kunth98} Kunth, D., et al. 1998, A\&A,
  334, 11
\bibitem[Larson(1974)]{larson} Larson, R. B., 1974, MNRAS, 169, 229
%\bibitem[Legrand et al(2001)]{leg01} Legrand, F., Tenorio-Tagle, G.,
%  Silich, S., Knuth, D., \& Cervi\~no, M. 2001, ApJ, 560, 630
\bibitem[Lehnert \& Heckman(1996)]{lehnny} Lehnert, M. D., \& Heckman,
  T. M. 1996, ApJ, 472, 546
%\bibitem[Lequeux et al.(1995)]{leq95} Lequeux, J., Knuth, D.,
%  Mas-Hesse, J., \& Sargent, W.  1995, A\&A, 301, 18
\bibitem[Leitherer et al.(1999)]{sb99} Leitherer, C., et
  al. 1999, ApJS, 123, 3 
\bibitem[Marlowe et al.(1995)]{am95} Marlowe, A. T., Heckman, T. M.,
  Wyse, R. F. G., \& Schommer, R. 1995, ApJ, 438, 563
\bibitem[Martin (1998)]{clm98}Martin, C. L. 1998, ApJ, 506, 222
\bibitem[Martin et al.(2002)]{clm02}Martin, C. L., Kobulnicky, H. A.,
  \& Heckman, T. M. 2002, ApJ, 574, 663 
\bibitem[Martin(2005)]{clm05} Martin, C. L. 2005, ApJ, 621, 227
\bibitem[McMillan et al.(1994)]{mcm94} McMillan, R., Ciardullo, R., \&
  Jacoby, G. H. 1994, AJ, 108, 1610
\bibitem[Mendez et al.(1999)]{mendez99} M\'endez, D. I., Esteban, C.,
  Filipovi\`c, M. D., Ehle, M., Haberl, F., Pietsch, W., \& Haynes,
  R. F. 1999, A\&A, 349, 801
%\bibitem[Moore(1970)]{moo70} Moore CE. 1970, Ionization Potentials and
%  Ionization Limits Derived from the Analysis of Optical
%  Spectra. Rep. No. NSRDS-NBS34. Washington, DC: US dept. Commerce 
\bibitem[Morton(1991)]{morton} Morton, D. C. 1991, ApJS, 77, 119
\bibitem[Mulder et al.(1995)]{foxy} Mulder, P. S., van Driel, W., \&
  Braine, J. 1995, A\&A, 300, 687
\bibitem[Murphy et al.(1996)]{murph96} Murphy, T. W., Jr., Armus, L.,
  Matthews, K., Soifer, B. T., Mazzarella, J. M., Shupe, D. L.,
  Strauss, M. A., \& Neugebauer, G. 1996, AJ, 111, 1025 
\bibitem[Ostlin et al.(1998)]{ost98} Ostlin, G., Bergvall, N., \&
  Roennback, J. 1998, A\&A, 335, 850
\bibitem[Ott, Walter \& Briggs(2005)]{ott} Ott, J, Walter, F., \& Briggs,
  Elias. 2005, MNRAS, 358, 1453
\bibitem[Pastoriza et al.(1993)]{past93} Pastoriza, M. G., Dottori,
  H. A., Terlevich, E., Terlevich, R., \& D\'iaz, A. I. 1993, MNRAS,
  260, 177
\bibitem[P\'erez-Montero \& D\'iaz(2003)]{pmd03} P\'erez-Montero,
  E. \& D\'iaz, A. I. 2004, MNRAS, 346, 105
\bibitem[Pettini et al.(2002)]{max02} Pettini, M., et al. 2002, ApJ,
  569, 742
%\bibitem[Phillips 1993]{phil93} Phillips, A. 1993, AJ, 105, 486
\bibitem[Robert et al.(1993)]{rob} Robert, C., Leitherer, C., \& Heckman,
  T. 1993, ApJ, 418, 749 
\bibitem[Rupke et al.(2002)]{rupke02} Rupke, D. S., Veilleux, S., \&
  Sanders, D. B. 2002, ApJ, 570,  588
\bibitem[Sauvage et al.(1997)]{stl97} Sauvage, M., Thuan, T. X., \&
  Lagage, P. O. 1997, A\&A, 325, 98 
\bibitem[Scannapieco \& Oh(2004)]{evanpeng} Scannapieco, E. D., \& Oh,
  S. P. 2004, ApJ, 608, 62
\bibitem[Schwartz \& Martin(2004)]{us} Schwartz, C. M. \& Martin,
  C. L. 2004, ApJ, 610, 201 
%\bibitem[Schwartz \& Martin(2006)]{us05} Schwartz, C. M. \& Martin,
%  C. L. 2006, in prep.
%\bibitem[Schwartz(2005)]{phd} Schwartz., C. M. 2005, Ph.D. Thesis, in prep.
\bibitem[Shapley et al.(2003)]{alice03} Shapley, A. E., Steidel,
  C. C., Pettini, M., \& Adelberger, K. L. 2003, ApJ, 588, 65
%\bibitem[Silich \& Tenorio-Tagle 1998]{st98} Silich, S. A., \&
%  Tenorio-Tagle, G. 1998, MNRAS, 299, 249
%\bibitem[Spitzer 1978]{spitz78} Spitzer, L. 1978, New York:
%  Wiley-Interscience, 1978 
%\bibitem[Spitzer \& Fitzpatrick(1995)]{sf95} Spitzer, L. \&
%  Fitzpatric, E. L. 1995, ApJ, 445, 196
\bibitem[Stil \& Israel2002]{2002A&A...392..473S} Stil, J. M., \&
  Israel, F. P. 2002, \aap, 392, 473
%\bibitem[Sutherland \& Dopita(1993)]{sd93} Sutherland, R. S. \&
%  Dopita, M. A. 1993, ApJS, 88, 253
\bibitem[Swinbank et al.(2004)]{swin04} Swinbank, A. M., Smail, I.,
  Chapman, S. C., Blain, A. W., Ivison, R. J., \& Keel, W. C. 2004,
  ApJ, 617, 64  
\bibitem[Thuan et al.(2004)]{tx04} Thuan, T. X., Hibbard, J. E., \&
  L\'evrier, F. 2004, ApJ, 128, 617
%\bibitem[Tremonti et al.(2001)]{tremonti01} Tremonti, C. A., Calzetti,
%  D., Leitherer, C., \& Heckman, T.M. 2001, ApJ, 555, 322
%\bibitem[Tully et al.(1981)]{tully81} Tully, R. B., Boesgaard, A. M.,
%  Dyck, H. M., \& Schempp, W. V. 1981, ApJ, 246, 38
\bibitem[Vacca \& Conti(1992)]{cow92} Vacca, W. D., \& Conti,
  P. S. 1992, \apj, 401, 543   
\bibitem[Viallefond \& Thuan(1983)]{vt83} Viallefond, F. \& Thuan,
  T. X. 1983, ApJ, 269, 444
\bibitem[V\'azquez et al.(2004)]{vaz04} V\'azquez, G. A. et al. 2004,
  ApJ, 600, 162
\bibitem[Vogt et al.(1994)]{hires} Vogt, S. S., et al. 1994,
  Proc. SPIE, 2198, 362 
\bibitem[Wilcots et al.(2001)]{wil01} Wilcots, E. M., Turnbull, M. C., \&
  Brinks, E. 2001, ApJ, 560, 110
\bibitem[van Woerden et al.(1993)]{vw93} van Woerden, H., van Driel,
  W., Braun, R., \& Rots, A. H. 1993, A\&A, 269, 15
\bibitem[Young et al.(1995)]{young95} Young, J. S., et al. 1995, ApJS,
  98, 219


\end{thebibliography}
\end{document}